\documentclass[twocolumn]{aastex63} 
\usepackage{amsmath}
\usepackage{mathtools}
\usepackage{lineno}
\submitjournal{PSJ}

\shorttitle{A Sublime Opportunity}
\shortauthors{Seligman, Kratter, Levine $\&$ Jedicke}
\graphicspath{{./}{figures/}}
\begin{document}

\title{A Sublime Opportunity: The Dynamics of Transitioning Cometary Bodies and the Feasibility of \textit{In Situ} Observations of The Evolution of Their Activity  }

\correspondingauthor{Darryl Z. Seligman}
\email{dzseligman@uchicago.edu}

\author[0000-0002-0726-6480]{Darryl Z. Seligman}
\affiliation{Dept. of the Geophysical Sciences, University of Chicago, Chicago, IL 60637}

\author[0000-0001-5253-1338]{ Kaitlin M. Kratter}
\affiliation{Steward Observatory, University of Arizona, 933 North Cherry Avenue, Tucson, AZ 85721, USA}

\author[0000-0002-1422-4430]{W. Garrett Levine}
\affil{Dept. of Astronomy, Yale University, 52 Hillhouse, New Haven, CT 06511, USA}

\author[0000-0001-7830-028X]{ Robert Jedicke}
\affiliation{University of Hawai‘i, Institute for Astronomy, 2680 Woodlawn Drive, Honolulu, HI 96822, USA}




\begin{abstract}
The compositional and morphological evolution of minor bodies in the Solar System is primarily driven by the evolution of their heliocentric distances, as the level of incident solar radiation regulates cometary activity. We investigate the  dynamical transfer of Centaurs into the inner Solar System, facilitated by mean motion resonances with Jupiter and Saturn. The recently discovered object, P/2019 LD2, will transition from the Centaur region to the inner Solar System  in 2063. In order to contextualize LD2, we perform N-body simulations of a population of Centaurs and JFCs. Objects between Jupiter and Saturn with Tisserand parameter $T_J\sim$3 are transferred onto orbits with perihelia $q<4$au within the next 1000 years with notably high efficiency. Our simulations show that there may be additional LD2-like objects transitioning into the inner Solar System in the near-term future, all of which have low $\Delta$V with respect to Jupiter. We  calculate the distribution of orbital elements resulting from a single Jovian encounter and show that objects with initial perihelia close to Jupiter are efficiently scattered to $q<4$au. Moreover, approximately $55\%$ of the transitioning objects in our simulated population experience at least 1 Jovian encounter prior to reaching $q<4$au. We demonstrate that a spacecraft stationed near Jupiter would be well-positioned to rendezvous, orbit match, and accompany LD2 into the inner Solar System, providing an opportunity to observe the onset of intense activity in a pristine comet \textit{in situ}. Finally, we discuss the prospect of identifying additional targets for similar measurements with forthcoming observational facilities.
\end{abstract}



\section{Introduction} \label{sec:intro}

The source of comets in the Solar System has been a long-standing subject of inquiry. Herschel and Laplace contemporaneously presented the idea that comets originated from outside of the Solar System \citep{Herschel1812, Herschel1812b,Laplace1814,Heidarzadeh2008}. \citet{Oort1950} explained the near-parabolical Long Period Comets (LPCs) by postulating the existence of the now-eponymous, spherical cloud of objects with isotropic inclinations at several $10^4$ au. However, the Oort cloud could not explain the strong tendency for Short Period Comets (SPCs), with periods, $P<200$yr, to lie near the ecliptic plane \citep{Everhart1972,Vaghi1973,Joss1973,Delsemme1973,Prialnik2020}.  Progress on the origin of the SPCs was made after the discovery of Pluto, when \citet{Leonard1930} hypothesized the existence of an ``Ultra-Neptunian'' population of planets beyond Pluto, an idea that was subsequently investigated by  many  authors \citep{Edgeworth1943,Edgeworth1949,Kuiper1951,Cameron1962,Whipple1964,Fernandex1980MNRAS,Duncan1988,Quinn1990}. This trans-Neptunian population was  confirmed with the historic detection of the first Kuiper Belt Objects (KBOs) by \citet{Jewitt1993}.  Since then, thousands of additional objects have been discovered from systematic campaigns such as the \textit{Deep Ecliptic Survey} \citep{Elliot2005} and the Outer Solar System Origins Survey (OSSOS) \citep{Volk2016,Shankman2016,Shankman2017,Bannister2018OSSOS}.

Moving closer to the Sun, the elusive Centaurs are commonly-defined as minor bodies  with perihelia outside of Jupiter ($q > 5.2$ au) and semi-major axes inside of Neptune ($a < 30.1$ au) \citep{Gladman2008}. The first identified member was 1977 UB, later renamed to (2060) Chiron \citep{Kowal1977}. Chiron was precovered in photographic plates from 1895 and 1941 \citep{Davies2001}. A second Centaur was discovered \citep{Scotti1992}, 1992 AD, which was then numbered and named (5145) Pholus. For a more detailed review of Centaur science, we guide the reader to \citet{Jewitt1999,Luu2002,Davies2004,Jewitt2008,Nesvorny2018,Ferrari2018,Peixinho2020}.

The region beyond Jupiter's orbit is too cold for substantial H$_2$O ice sublimation. Therefore, the Centaurs are a mix of inactive asteroidal and cometary bodies with activity driven by more volatile substances, such as CO and CO$_2$ \citep{Barnun1988,Prialnik1990,Barnun1998,Womack2017}. For example, \citet{Meech1990} presented evidence of cometary activity in Chiron. Moreover, 2000 EC98, since re-classified as periodic comet 174P/Echeclus, is well-known for its massive outburst in 2005 \citep{Choi2006,Bauer2008} and smaller ones in 2011, 2016, and 2017 \citep{Kareta2019,Jaeger2011,James2018}. \citet{Jewitt2009} found that 9 of a sample of 23 observed Centaurs displayed activity consistent with the release of trapped gases as amorphous ice was converted to the crystalline form. CO and  CO$_2$ have been detected in distant comets, such as in a sample of 163 comets from the Wide-Field Infrared Survey Explorer (WISE) \citep{Bauer2015}.

The complex dynamical evolution of Centaurs is dominated by gravitational perturbations from the giant planets. \citet{Hahn1990} presented numerical simulations demonstrating that the orbit of Chiron was chaotic on a $\sim 0.5$ Myr timescale and that the object was likely to become an SPC in the future. This study showed that Chiron's orbital evolution could be representative of the broader Centaur and SPC populations, driven by mean motion resonances (MMRs) with giant planets in analogy to the asteroid belt's Kirkwood gaps \citep{Wisdom1983,Morbidelli2002}.

\citet{Levison1997} presented numerical simulations of Centaurs migrating from the Kuiper belt, and found that about $\sim30\%$ of these objects reached $q<2.5$au. \citet{Tiscareno2003} presented long-term dynamical simulations of the 53 known Centaurs in 2003, and found that approximately $2/3$ of Centaurs were scattered into the Oort cloud with nearly all others becoming SPCs. They found that the median dynamical lifetime of a Centaur was 9 Myr, with a large scatter between 1-100 Myr, and that Centaurs spent most of this time on orbits with eccentricities between 0.2 and 0.6 and perihelia between 12-30 au. \citet{DiSisto2007} found that a small fraction of Centaurs impacted a giant planet or  became scattered disk objects (SDOs), but did not become cold classic KBOs. 
\citet{Bailey2009} identified two types of chaotic evolution for Centaurs, one  exhibiting random walks in the orbital evolution, and one whose evolution is dominated by intermittent resonance sticking, with stochastic jumps between MMRs. \citet{Nesvoryn2017} performed simulations of Centaur evolution over 4.5 Gyr timescales, including the hypothetical Planet Nine \citep{Batygin2015}, and reproduced the distribution of SPCs. \citet{Fernandez2018}  found that the median lifetime of inactive Centaurs was $\sim2\times$ longer than that for active Centaurs, implying a connection between activity and residence time of Centaurs. This was corroborated by the fact that the high inclination and retrograde Centaurs are all inactive and have the longest lifetimes. They found that active Centaurs, unlike the inactive Centaurs, experienced close approaches to the Sun in their recent lifetime.

In 2019, the Asteroid Terrestrial-impact Last Alert System (ATLAS) discovered the object, P/2019 LD2, which is  an active Centaur that is likely to become a JFC in the current century \citep{Kareta2020}. \citet{Steckloff2020} demonstrated that this transition will happen after a close approach to Jupiter in 2063. They  performed simulations of LD2's history over the last 3,000 years and found that it was unlikely to have spent time in the inner Solar System, implying that its future transition  will be its first close encounter with the Sun. \citet{Hsieh2021} showed that LD2 only reached its current orbit in July of 2018. Recently, \citet{Sarid2019} identified a ``Dynamical Gateway", in which $\sim1/2$ of the Centaurs that became JFCs briefly occupied  before transitioning.  This region is characterized by objects on nearly circular orbits  just outside of the orbit of Jupiter, and represents a surprisingly small fractional area of parameter space relative to the fraction of JFCs that pass through it. \citet{Steckloff2020} demonstrated that LD2 was a recent occupant of the Gateway. The well studied active Centaur 29P/SW1 \citep{Senay1994,Crovisier1995,Gunnarsson2008,Paganini2013,Schambeau2018,Fernandez2018,Wierzchos2020} is currently in the Gateway, and may transition to the inner Solar System in the next $\sim10^4$ years,  like LD2 \citep{Sarid2019}. The discovery of LD2 presents an opportunity to closely observe the transition from an active Centaur to a JFC with an orbit matching spacecraft.  %They showed that LD2 will become a Centaur in 2028 February, before{ eventually transitioning to becoming Jupiter-family comet in 2063 February. 

Over the past few decades, a remarkable array of missions have visited small bodies in both the inner and outer Solar System. Spacecraft have investigated a diversity of asteroids \citep{chen1997near, russell2015dawn}, comets \citep{Tsurutani1986, Nelson2004, Glassmeier2007}, and Kuiper Belt Objects \citep{stern2019arrokoth} and some have even returned samples  to Earth \citep{Tsou2004,Lauretta2012, watanabe2017hayabusa2}. Nonetheless, no spacecraft has ever visited a Centaur. Two recent candidates for NASA's Discovery program -- \textit{Centaurus} 
\citep{singer2019centaurus} and \textit{Chimera} \citep{harris2019chimera} --  were proposed to visit and study primarily Chiron and 29P/SW1, respectively, but were not selected for further development. In this study, we examine the feasibility of a mission that would rendezvous with LD2 as it begins its journey into the inner Solar System. This park and wait approach is similar to that of the upcoming ESA mission \textit{Comet Interceptor} \citep{jones2019}, which will rendezvous with an as of yet unidentified LPC or perhaps an interstellar object.

This paper is organized as follows. In \S \ref{sec:dynamics}, we highlight the overlapping MMRs with Jupiter and Saturn in the Gateway that enable the transfer between Centaurs and JFCs, in order to contextualize LD2 within its host population. In \S \ref{sec:simulations} and \ref{sec:fullcentaur}, we perform N-body simulations to identify initial conditions and estimate the number of objects that will transition into the region where $q<4$au in the next 1000 years, in order to assess the possibility that there will be additional targets for a rendezvous mission. In \S\ref{sec:scattering}, we calculate the distribution of orbital elements following a close encounter with Jupiter, and identify  regions of orbital space where objects can be scattered onto orbits with $q<4$au via a single close encounter with Jupiter. We show that LD2 is representative of its class in terms of its dynamical transfer, orbital evolution, and feasibility for a rendezvous mission. In \S \ref{sec:missions}, we demonstrate the feasibility of an orbit matching rendezvous with LD2 after the 2063 Jovian encounter. In \S\ref{sec:discussion}, we discuss the detection prospects for these objects with forthcoming observatories and conclude. 

\section{Dynamics of the Gateway Region}\label{sec:dynamics}
\begin{figure*}
\begin{center}
\includegraphics[scale=0.6,angle=0]{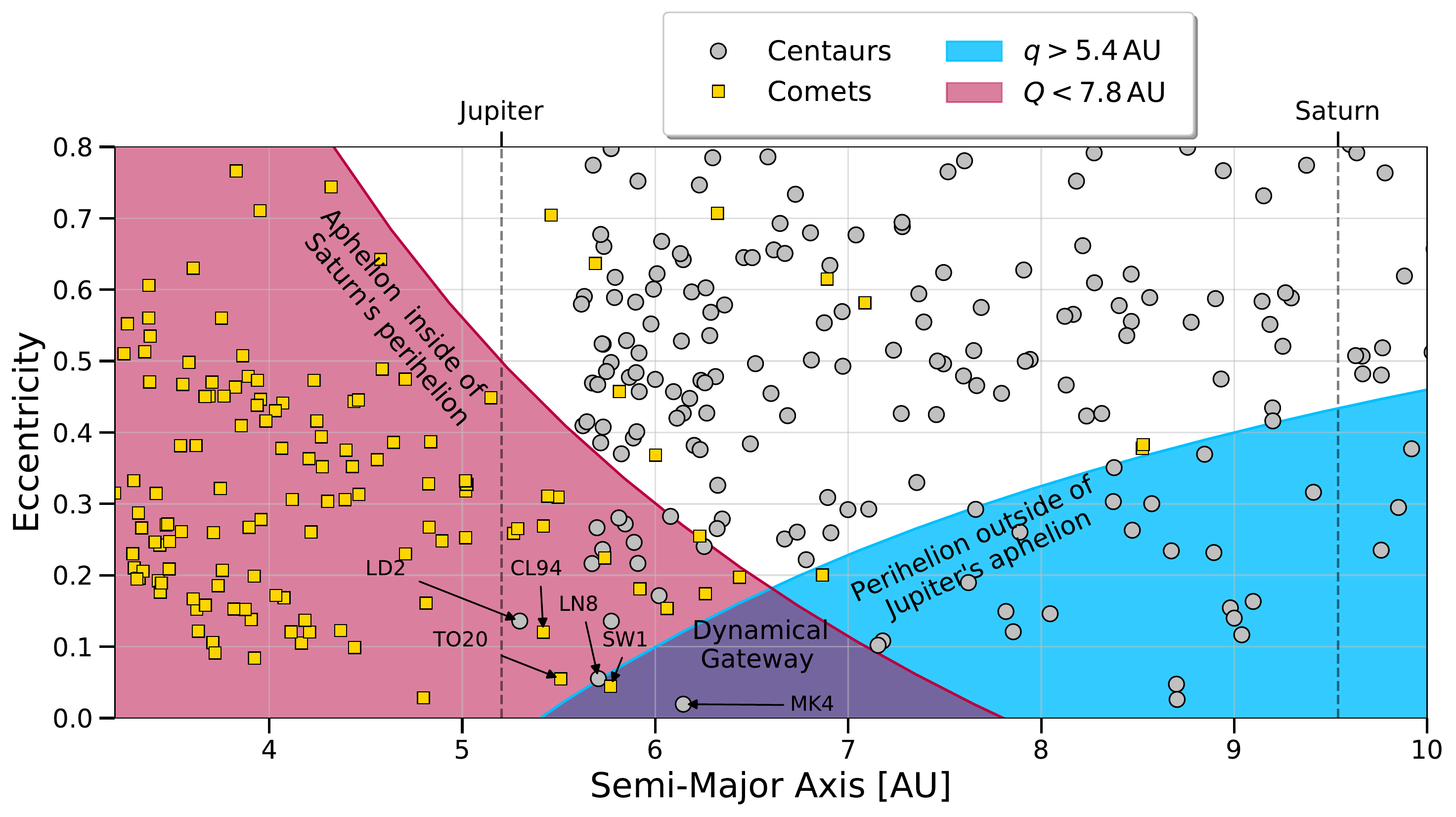} % this command will be ignored
\caption{The  Gateway depicted in eccentricity verses semi-major axis space. The red shaded area indicates the region in which a body's aphelion is greater than 3 Hill radii from Saturn’s perihelion. The blue shaded area indicates the region where a body's perihelion is outside of Jupiter's aphelion. The region where both of these criteria are satisfied defines the  Gateway \citep{Sarid2019}, where $\sim1/2$ of the JFCs pass through on their journey into  the inner Solar System. The yellow squares and grey circles indicate currently known comets and Centaurs respectively, as labeled in the MPC database. The positions of Jupiter and Saturn and 29P/ SW1, P/2010 TO20, P/2008 CL94 and 2016 LN8, which were identified by \citet{Sarid2019} as current or recent occupants of the Gateway, are indicated. The location of P/2019 LD2 and MK4 are also indicated.}\label{Fig:gateway_region}
\end{center}
\end{figure*}
\begin{figure*}
\begin{center}
\includegraphics[scale=0.45,angle=0]{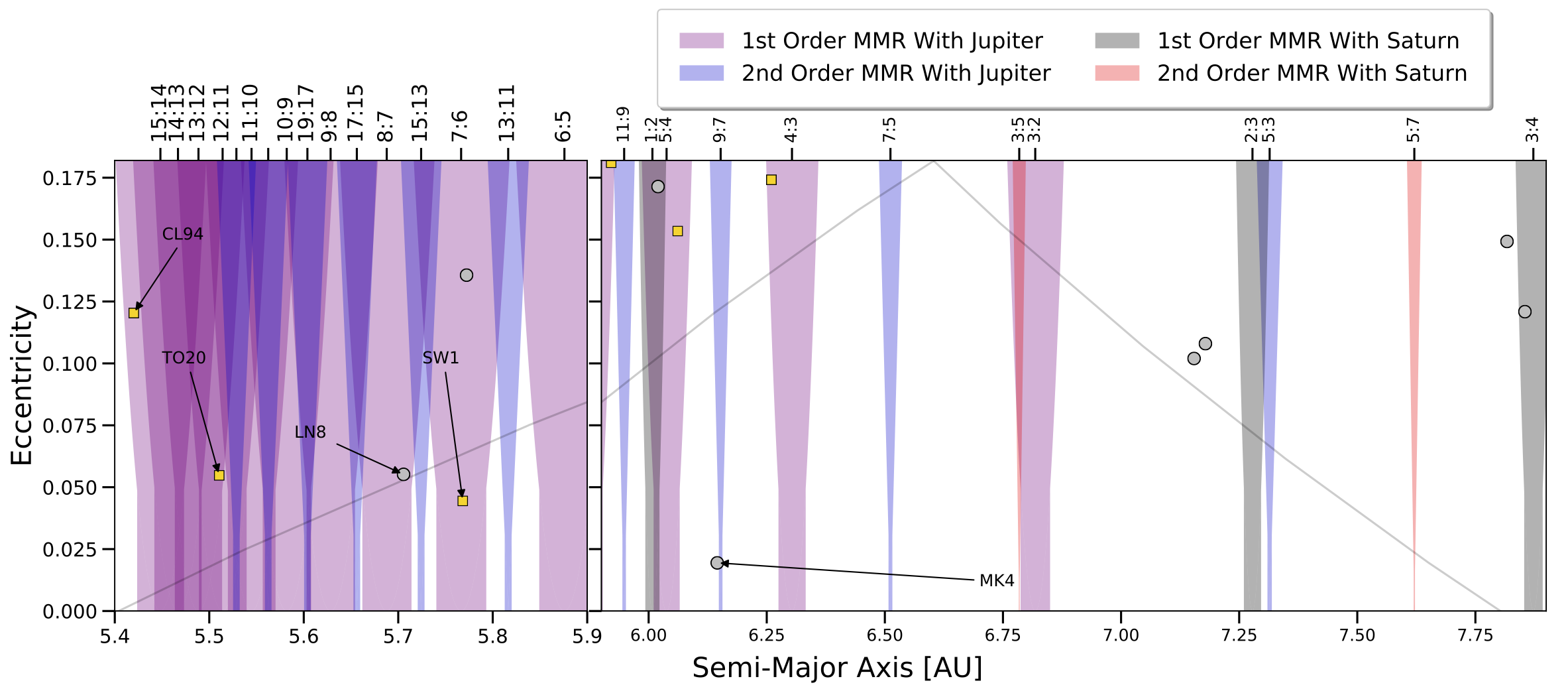} % this command will be ignored
\caption{The  locations and structure of 1st and 2nd order MMRs with Jupiter and Saturn in the Gateway. First and second order  MMRs with Jupiter and Saturn are shown in purple and blue, and grey and red, respectively . The 1:2 and 2:3 first order and 5:7 and 3:5 second order MMRs with Saturn are within the Gateway. The external first order resonances of degree higher than 3 and second order resonances of degree higher than 5 with Jupiter are within the Gateway. In the interior region of the Gateway, depicted in the left panel, the MMRs with Jupiter overlap and create a region in parameter space prone to chaos. The 5:4 and 1:2, 3:2 and 3:5 and 2:3 and 5:3 MMRs with Jupiter and Saturn also overlap.}\label{Fig:MMR_saturn_jupiter}
\end{center}
\end{figure*}
%\subsection{The Structure of the Dynamical Gateway}

The  Gateway,  introduced by \citet{Sarid2019}, is a region in semi-major axis and eccentricity space from which Centaurs sourced from the SDOs are scattered into the inner Solar System. These authors presented numerical simulations that tracked a large number of TNO test particles through the Centaur population and into the JFC region.  About $1/2$ of the JFCs produced in their simulations occupied the Gateway prior to transitioning. Moreover, objects only remained in this region for $\sim100-1000$ years before becoming JFCs, as outlined in Table 1 of their paper.   They defined the Gateway as orbits with aphelion, $Q<7.8$au, and perihelion, $q>5.4$au. These limits demand that the orbits do not cross Jupiter's, and that the aphelion is greater than 3 Saturnian Hill radii away from Saturn's  perihelion. 

In Figure \ref{Fig:gateway_region}, we show the location of the Gateway and of Solar System objects currently in the Minor Planet Center database, including the objects P/2019 LD2, 29P/ SW1, P/2010 TO20, P/2008 CL94 and 2016 LN8 which were identified as current or recent occupants of the Gateway \citep{Sarid2019,Kareta2020,Hsieh2021,Steckloff2020}. Since the publication of these papers, the Centaur 2020 MK4 was detected\footnote{\href{https://www.minorplanetcenter.net/mpec/K20/K20N36.html}{Link}}, which  resides  within the limits of the  Gateway as well \citep{delafuente2021}. It is somewhat striking that there are so few currently known objects in the Gateway region, compared to the observed Centaurs and comets in the MPC database. There is a selection effect, because objects with higher eccentricity are easier to detect since they attain lower perihelia. The fact that there are so few objects currently detected in this region could be attributed to observational selection effects and/or the low median residency time \citep{Sarid2019,Steckloff2020}.

%In Figure \ref{Fig:gateway_region}, we show this Gateway region in eccentricity verses semi-major axis. The blue and red shaded regions correspond to the constraint that $q>5.4$ au and $Q<7.8$ au respectively. The overlapping region that defines the Gateway is shown in purple, and spans from $a\sim5.4$ to $\sim7.8$ au and  eccentricity up to $e\sim 0.2$.  The points indicate the position of comets and centaurs currently  in the Minor Planet Center database. The yellow points are all of the objects in the MPC file for comets, and the grey poinst are all of the objects in the catalogue of distant objects that satisfy $a_J<a<a_N$.   We have indicated the location of the objects P/2019 LD2, 29P/ SW1, P/2010 TO20, P/2008 CL94 and 2016 LN8 which were identified   as current or recent occupants of the Gateway \citep{Sarid2019,Kareta2020,Hsieh2021,Steckloff2020}. Since the publication of these papers, the Centaur 2020 MK4 was detected\footnote{https://www.minorplanetcenter.net/mpec/K20/K20N36.html}, which  resides  within the limits of the dynamical Gateway as well \citep{delafuente2021}.

%\subsection{Mean Motion Resonances with Jupiter and Saturn as an Explanation for the Gateway Region}

Since Gateway orbits are characterized by low eccentricities and semi-major axes close to Jupiter, a natural explanation for the transient nature of objects in this region is from gravitational interactions with Jupiter and Saturn.  We investigate the effect of first and second order MMRs with Jupiter and Saturn on objects specifically in this region, as was done for the entire Centaurs region in \citet{Bailey2009,Tiscareno2003}. We do this in order to contextualize LD2 within its host population, and investigate the mechanism that generates objects like LD2 with $q<4$au in short timescales.  For the configuration of the gravitational interactions of the Sun, Jupiter and a Gateway object,  the circular planar restricted three body problem represents a reasonable approximation. \citet{Wisdom1980} derived the resonance overlap criteria for MMRs, which is  commonly referred to as the $\mu^{2/7}$ scaling law,  where $\mu$ is the mass of Jupiter divided by the mass of the Sun. The location of an interior MMR of order q is of the form $p+q:p$, where p and q are integers, and is defined by orbits with $(p+q)n_2=pn_1$, where $n_1$ and $n_2$ are the mean motions of the inner and outer body. From Section 6.1 in \citet{Malhotra1998}, neighboring MMRs where $q=1$, of the form $p+1:p$ will begin to overlap for integers p such that $p^{-1}<2.1\mu^{2/7}$. Moreover, the approximate widths for  resonance angle librations are,

\begin{equation}
    \Delta n/n \sim \mu^{2/3}\, ,
\end{equation}
\noindent for close to circular orbits, and
\begin{equation}
    \Delta n/n \sim (e \mu)^{1/2}\, ,
\end{equation}
for first order resonances. 

In Figure \ref{Fig:MMR_saturn_jupiter}, we show the nominal locations of the first and second order  MMRs with Jupiter and Saturn of the form $p+1:p$, $p+2:p$, $p:p+1$, $p:p+2$ in eccentricity and semi-major axis space. For each MMR, the shaded regions show the  librations  of eccentricity and semi-major axis within the resonance.  The range of semi-major axis variations from circular resonances are filled in the same color for each MMR, up until eccentricities of 0.025. The two constraints on perihelion and aphelion representing the Gateway region are shown in solid lines. MMRs with Uranus and Neptune are negligible for this region of orbital space. In the  regions close to Jupiter,  the overlap of these MMRs could be responsible for the ejection of bodies. There is also a dense region of resonance overlap at low semi-major axis and eccentricity above the Gateway, in the upper left hand side of the left panel of Figure  \ref{Fig:MMR_saturn_jupiter}.%, which is consistent with the fact that  these objects are perturbed on short timescales. 

%\subsection{Numerical Simulations  of Transitioning Centaurs}
\section{Numerical Simulations of Gateway Objects}\label{sec:simulations}
%\subsection{Lifetime of Centaurs in the Gateway}
In this section, we investigate the dynamics of objects that begin in the Gateway and are transferred into the inner Solar System. The purpose of these simulations is to explore the dynamics of LD2 as they pertain to the broader population, and estimate if there will be more transitioning targets for a rendezvous mission like the one proposed in \S \ref{sec:missions}. In Subsection \ref{subsec:definitions}, we review literature definitions of cometary minor bodies. In Subsection \ref{subsec:ICs}, we describe the initial conditions of our numerical simulations. In Subsection \ref{subsec:MMR}, we present the initial conditions and orbital evolution of objects that are transferred  into the inner Solar System in our simulations. In Subsection \ref{subsec:transferrates}, we compare our  simulations to the observed injection rates of JFCs to estimate the frequency with which LD2-like objects will transition into the inner Solar System. 

\begin{figure}
\begin{center}
\includegraphics[scale=0.4,angle=0]{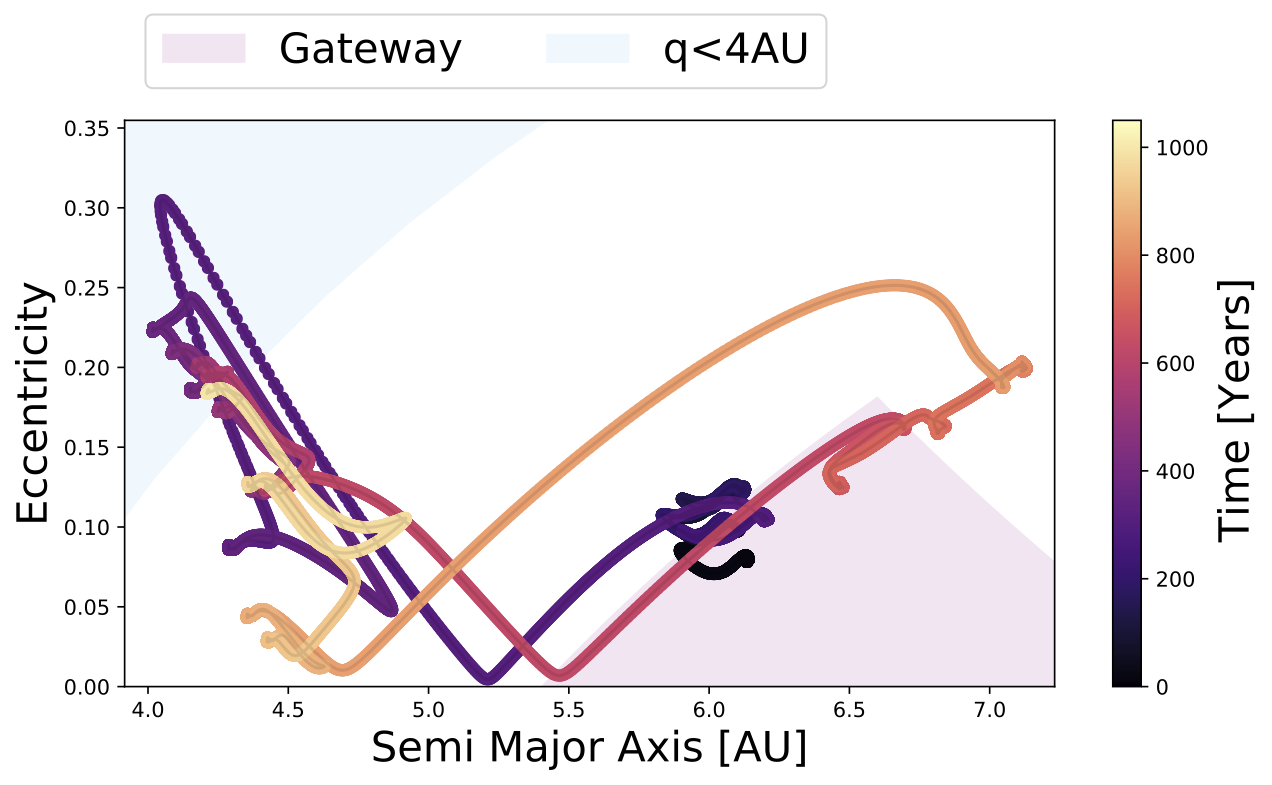}
\includegraphics[scale=0.4,angle=0]{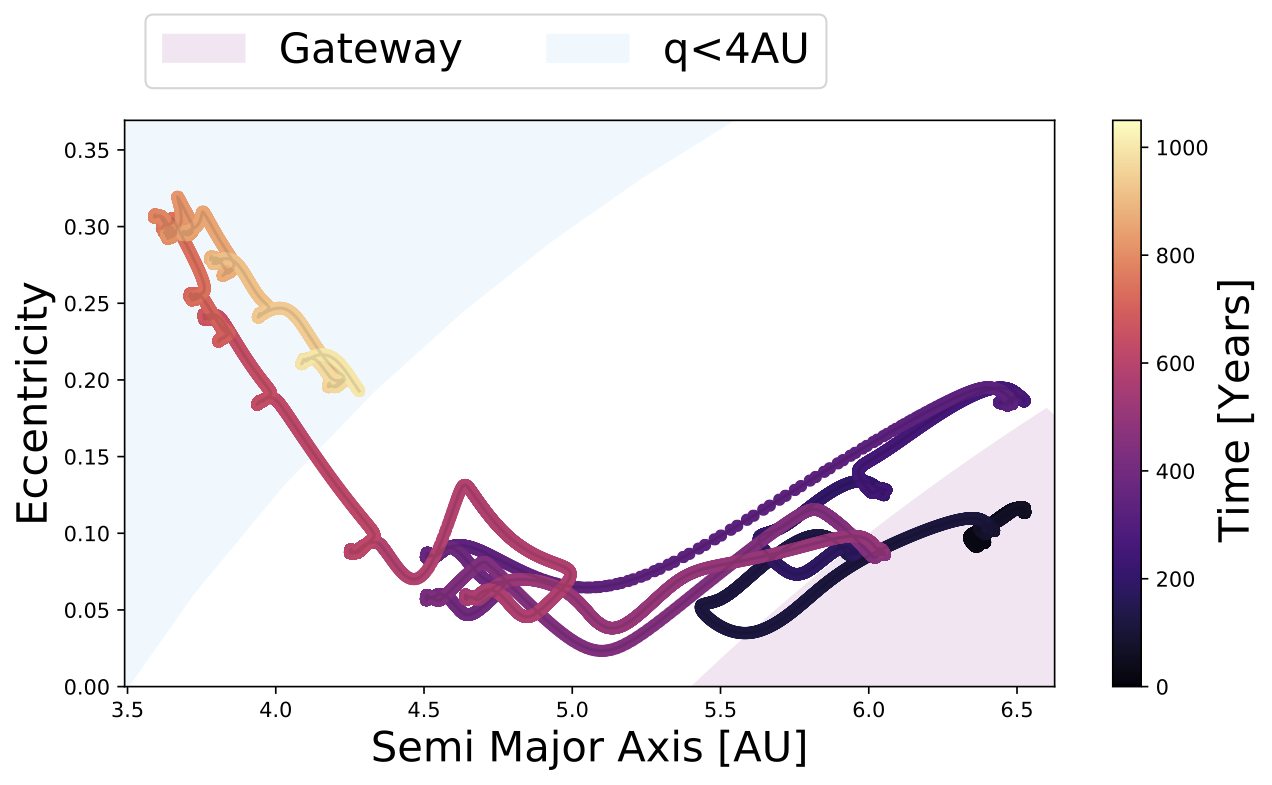}
\includegraphics[scale=0.4,angle=0]{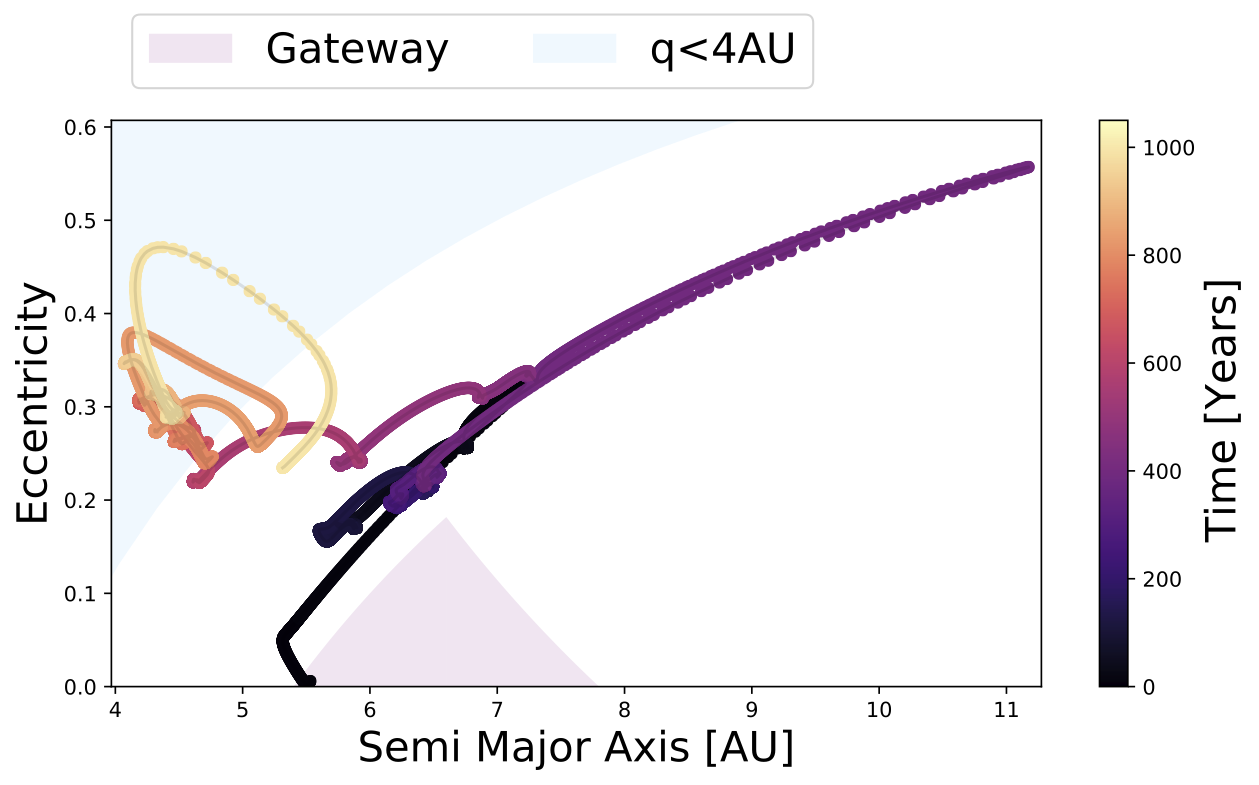}
\includegraphics[scale=0.4,angle=0]{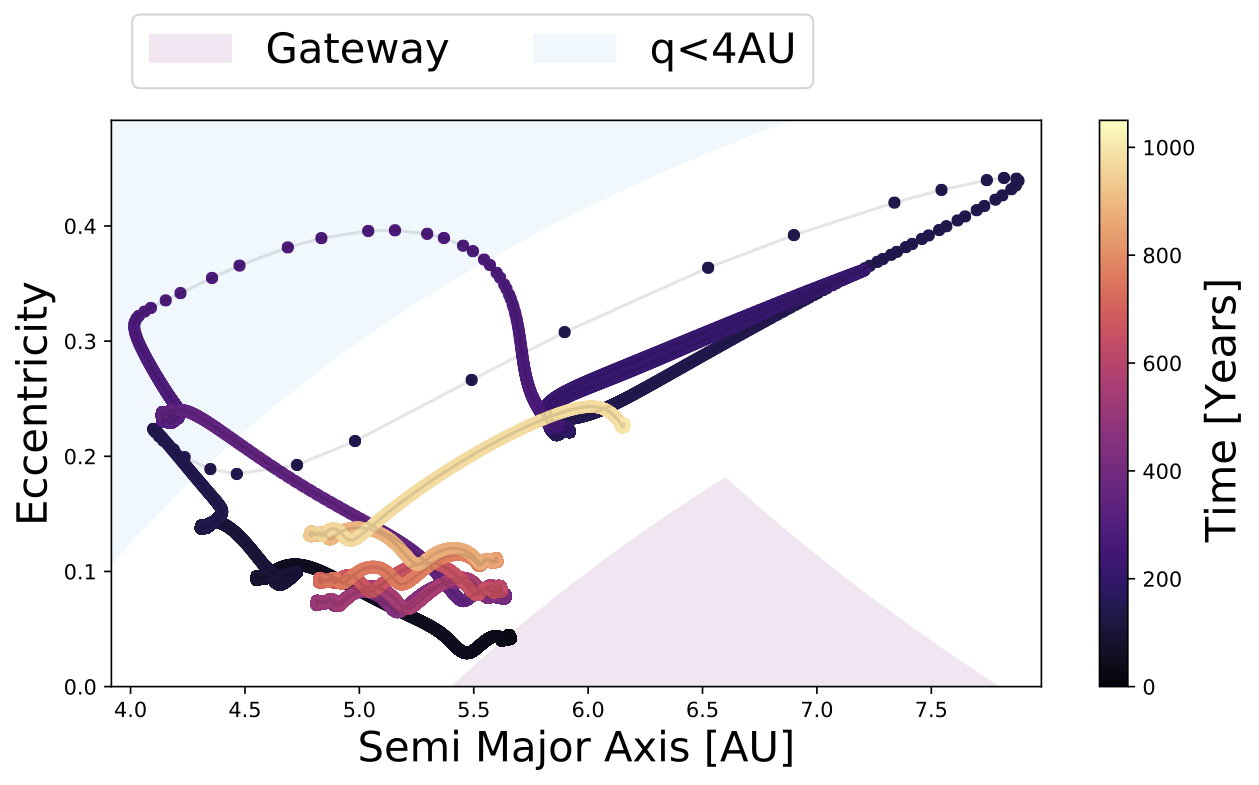}
\caption{Examples of the orbital evolution of objects that begin in the  Gateway and reach $q<4$au within the next 1000 years. The Gateway region  is shaded in purple, and the region where $q<4$au is shaded in light blue. The orbital evolution for each object is shown in colored points every $\sim .01$ years, and the color-scale corresponds to  time in the simulation.}\label{Fig:orbits_examples}
\end{center}
\end{figure}

\subsection{Literature Definitions and Population Estimates of Cometary Bodies Interior to Neptune}\label{subsec:definitions}

\begin{table*}[]
    \centering
    \begin{tabular}{c|c|c|c|c|c|c}
          Object & Definition &Number &Reference\\
        \hline
         SPC& $T_J>2$ and $a<a_N$ &$\sim1.2\times10^7$ &\citet{Levison1997}  \\
         Encke type comets& $T_J>3$ and $a<a_J$ &&\citet{Levison1996}  \\
         Chiron-like  comets& $T_J>3$ and $a>a_J$ &&\citet{Levison1996}  \\
         JFC& $3>T_J>2$ &&\citet{Levison1996}  \\
         JFC& $q<5.2$au and $Q<7$au &&\citet{Sarid2019}  \\
         JFC& $2<T_J<3.1$ and $P<20$yr &$450\pm50$ with $R\gtrsim 1$km&\citet{Disisto2009}  \\
         OJFC& $q < 2$ au and $H<9$&$\sim108 $ &\citet{Levison1997}  \\
         OJFC& $q<2.5$au and $H<10.8$ &$\sim 294^{+556}_{-235} $ with $R\gtrsim 1.15$km&\citet{Brasser2015}  \\
         OJFC&$2 < T_J < 3$ and $q < 2.5$ au  &$\sim355 $ with $R\gtrsim 1$km&\citet{Roberts2021}  \\
         GPC& $5.2<q<30$au &$\sim2.23\times10^8$  with $R\gtrsim 1$km&\citet{Disisto2020}  \\
         NC& $q>5.204$au and $5.6<Q<9.583$au &$262\pm55$  with $R\gtrsim 1$km&\citet{Roberts2021}  \\
         Centaur& $5.2<a,q<30$au &&\citet{Jewitt2009}  \\
         Centaur& $5.2<a,q<30$au &$\lesssim4\times10^7$ with $R\gtrsim 1$km&\citet{Jedicke1997}  \\
         Centaur& $5.2<a<30$au &$\sim10^6$  with $R \gtrsim1$km&\citet{Disisto2020}  \\
         Centaur& $5.2<a,q<30$au &$\sim6.5\times10^6$  with $R\gtrsim 1$km&\citet{Sarid2019}  \\
         Centaur& $q>7.5$ au and $a < a_N$  &$\sim6\times10^5$  with $R\gtrsim 1$km&\citet{Nesvorny2019}  \\
         Centaur& $a_J<a,q<a_N$ &$\sim3.6 \times10^5$ with $R\gtrsim 1$km&\citet{Roberts2021}  \\
    \end{tabular}
    \caption{Various proposed definitions of Short Period Comets (SPCs), Encke type comets, Chiron-like comets, Jupiter Family Comets (JFCs), Observable Jupiter Family Comets (OJFCs), Near Centaurs (NCs), Giant Planet Crossers (GPCs) and Centaurs. Population number estimates are shown with size-frequency distributions extended to $R\gtrsim 1$km, where applicable.  }
    \label{tab:definitions}
\end{table*}

Before describing our simulations, it is useful to review the multiple definitions that cometary minor bodies in the Solar System have been given and the number estimates of these populations.  \citet{Levison1996} and \citet{Levison1997} defined SPCs as objects inside the trans-Neptunian region with a Tisserand parameter with respect to Jupiter, $T_J$, that satisfies,
\begin{equation}\label{eq:tisserand}
    T_J=a_J/a+2\cos(i)\sqrt{a/a_j(1-e^2)}>2\,,
\end{equation}
 where $a_J$ is Jupiter's semi-major axis. They divided the SPCs into three groups:

\begin{itemize}
    \item Jupiter Family Comets with $2<T_J<3$, which can experience low velocity encounters with Jupiter and whose dynamics are dominated by the giant planet. 
    \item Encke type comets, with $T_J>3$ that cannot cross Jupiter's orbit, and  $a<a_J$.
    \item Chiron-like comets with  $T_J>3$ and   $a>a_J$.
\end{itemize} 

 \citet{Disisto2020} adopted the definition for JFC as an object with $q<5.2$au, referring to the population of bodies with $5.2<q<30$ au as giant planet crossers (GPCs) while Centaurs were those with $5.2<a<30$ AU. \citet{Roberts2021} defined the Near Centaurs (NCs) as objects with  $q > 5.204$ au and aphelion $5.6 < Q < 9.583$ au.  \citet{Jewitt2009} proposed the definition for Centaurs as  objects with $q$ and $a$ between the semi-major axes of Jupiter and Neptune that are not in 1:1 MMRs with any giant planet. We adopt this definition of a Centaur for the remainder of this paper. 

Estimating the number of Centaurs has also been the focus of a large number of studies. \citet{Jedicke1997} determined the efficiency of the Spacewatch system as a function of an object's apparent visual magnitude and rate of motion. With the then known population of discovered and a synthesized population of Centaurs, they estimated that there were fewer than $\sim2000$ Centaurs in the Solar System, and  $\sim 3$ objects with diameters of $\sim200$km or larger, comparable to Chiron. If we extrapolate the size-frequency distribution in this paper to $R\gtrsim 1$km, it would suggest there are $\lesssim4\times10^7$ Centaurs.  \citet{DiSisto2007} (updated in \citet{Disisto2020}) generated a synthetic population of Centaurs that were sourced solely from the SDOs, and found that this source could generate $\sim2.8\times10^8$ GPCs with $R\gtrsim 1$km. They found good agreement between the magnitude corrected synthetic survey and the population of known Centaurs at the time. However, they admitted that there were very likely other source regions for Centaurs that could change the underlying orbital distribution. For example, \citet{Kazantsev2021} demonstrated that the main asteroid belt  could also source objects into the Centaur region via MMRs with Jupiter. Several authors have presented additional estimates of the size and definitions of the Centaur population \citep{Sarid2019,Nesvorny2019,Roberts2021}, which we summarize in Table \ref{tab:definitions}. Not including the definitions based on the Tisserand parameter, almost all of the definitions for a given class of minor bodies broadly overlap.

%\citet{Sarid2019} estimated that there were $\sim6.5\times10^6$ centaurs with $R>1km$, defined by $q>5.2$ au and $Q < 30.1$ au. \citet{Nesvorny2019} estimated that there were $\sim6\times10^5$ centaurs with $R>1km$, defined by $q>7.5$ au and $a < a_N$, the semi-major axis of Neptune. \citet{Disisto2020} estimated that there were $\sim10^6$ centaurs with $R>1km$, defined by $5.2<a<30$ au. \citet{Roberts2021} estimated that there were $\sim3.6 \times10^5$ centaurs with $R>1km$, defined by $a_J<a,q<a_N$.

There have also been extensive efforts to estimate the size of the population of SPCs, which is an important normalization factor for estimating the population of Centaurs and Kuiper belt objects from numerical simulations. \citet{Levison1997}  found that JFCs had average dynamical lifetimes of $12,000$ years and estimated there were $1.2 \times10^7$ SPCs. This short lifetime suggested a constant injection rate from the Centaurs \citep{Dones2015}. \citet{Disisto2009} examined the distribution of JFCs that were simulated from the scattered disk, using the definition for JFC as an object with Tisserand parameter $2<T_J<3.1$. They included detailed cometary fading, non-gravitational forces, sublimation and splitting models to reproduce the observed population of JFCs. They estimated that there were $450\pm50$ JFCs with radii $R\gtrsim 1$km. They also estimated that the population of non-JFCs with Jupiter crossing orbits, objects that satisfy the Tisserand parameter requirement but do not reach orbital periods less than 20 years or have semi major axis  $a>7.37$ AU, was roughly 4 times larger, $\sim2250\pm250$ with $R\gtrsim 1$km. 

The observable JFCs (OJFCs) are defined as bodies with $2 < T_J < 3$ and $q < 2.5$ au \citep{Levison1997,Rickman2017}, and are used as a nominally complete sample of objects. The current number of known OJFCs quoted in \citet{Roberts2021} is  355, which they assumed to represent a complete and steady state population with radius $R \gtrsim 1$km, since it matched previous population estimates and was used to validate their Centaur population. However, they noted that ``the current population of OJFCs is very unlikely to be complete, and even if it was, we do not know the diameters of most of the objects.'' These definitions and number estimates are summarized in Table \ref{tab:definitions}.

\subsection{Initial Conditions and Numerical Simulation Details}\label{subsec:ICs}
We wish to  identify the initial conditions for objects in the Gateway that reach  $q<4$au within the next 1000 years, that could serve as additional targets for a rendezvous mission.  In order to sample this population, we perform numerical N-body  simulations of $\sim 100,000$ test particles that begin within the Gateway with the REBOUND N-body code \citep{rebound} along with the terrestrial and giant planets. We use a time step that is $1/60$ of Mercury's period. The simulations were integrated using the hybrid symplectic MERCURIUS integrator \citep{reboundmercurius}, which is appropriate for close encounters.  We draw  the semi-major axis and eccentricity for each test particle from uniform distributions within the Gateway  in order to densely sample this region. We  draw inclinations from a normal distribution centered at 0 with standard deviation of 30$^\circ$, and randomly draw longitude of periapse, $\omega$, and longitude of ascending node, $\Omega$. We integrate each test particles for 1000 years starting in 2021. 

It is important to note that the initial conditions drawn from uniform orbital elements were not chosen to represent the underlying distribution of orbital elements of the Centaurs and JFCs in the Solar System. While the underlying distribution is observationally unconstrained, many authors have performed extensive numerical simulations starting with objects in the trans-Neptunian region, tracking their evolution through the Centaur region into the JFC region, to estimate the orbital element distribution of the Centaur population \citep{DiSisto2007,Disisto2009,Disisto2020,Bailey2009,Tiscareno2003,Roberts2021,Sarid2019}. 

In order to be agnostic about the source population and be computationally efficient, these simulations are designed to investigate the dynamics of objects like LD2. Studying the evolution of objects like LD2 numerically should give us insight into the onset of intense sublimation, if LD2 is representative of its class. These simulations are simply designed to give us a representation of what the evolutionary dynamical pathways are for objects that become bright water driven comets. As we show in the next two subsections, it is likely that  the  region is not uniformly populated. In Subsection \ref{subsec:transferrates},  we  revise the estimates for the population based on the likely sampling depletion and sparse population  of this region.

\subsection{Objects that Reach the Inner Solar System}\label{subsec:MMR}
In order to explore the role of MMRs in the Gateway, we examined  the temporal evolution of our simulated objects. In Figure \ref{Fig:orbits_examples}, we show several examples of the orbital evolution of these objects.  While these plots do not show the full evolutionary picture of every object that transitions, they provide informative examples of the  dynamical pathways that can lead to the generation of an active comet in the inner Solar System. 
%The Gateway region itself is shaded in purple, and the region where $q<4$au is shade in light blue. The trajectory for each object is shown in colored points, where the color-scale corresponds to  time in the simulation.

\begin{figure}
\begin{center}
\includegraphics[scale=0.35,angle=0]{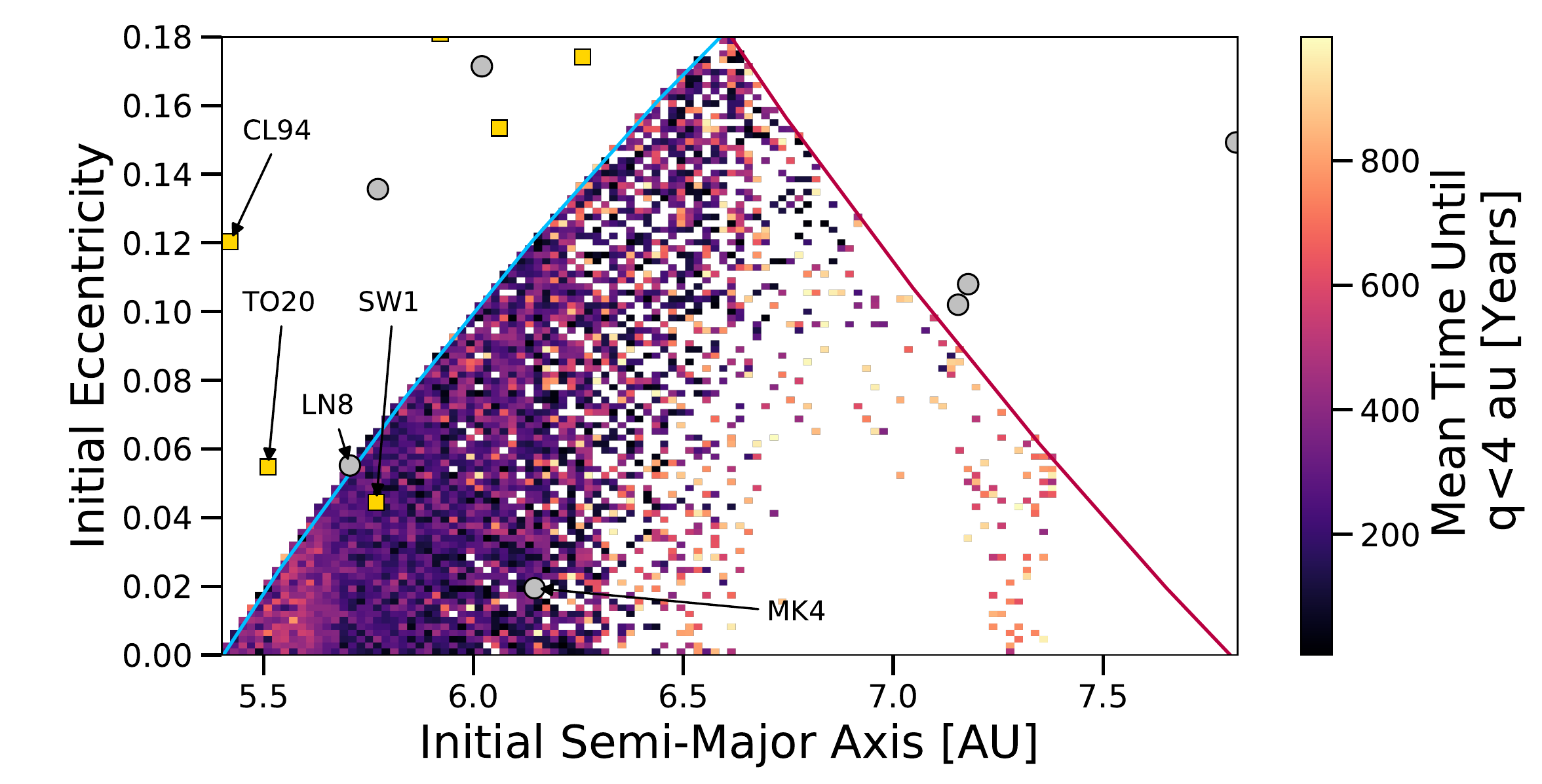}
\caption{The time until  objects  that start in the Gateway reach  $q<4$au. The color indicates the time elapsed before this occurs, averaged over all simulated objects. Currently known Centaurs and JFCs are indicated in grey and yellow points.}\label{Fig:lifetimes_Gateway}
\end{center}
\end{figure}

\begin{figure*}
\begin{center}
\includegraphics[scale=0.45,angle=0]{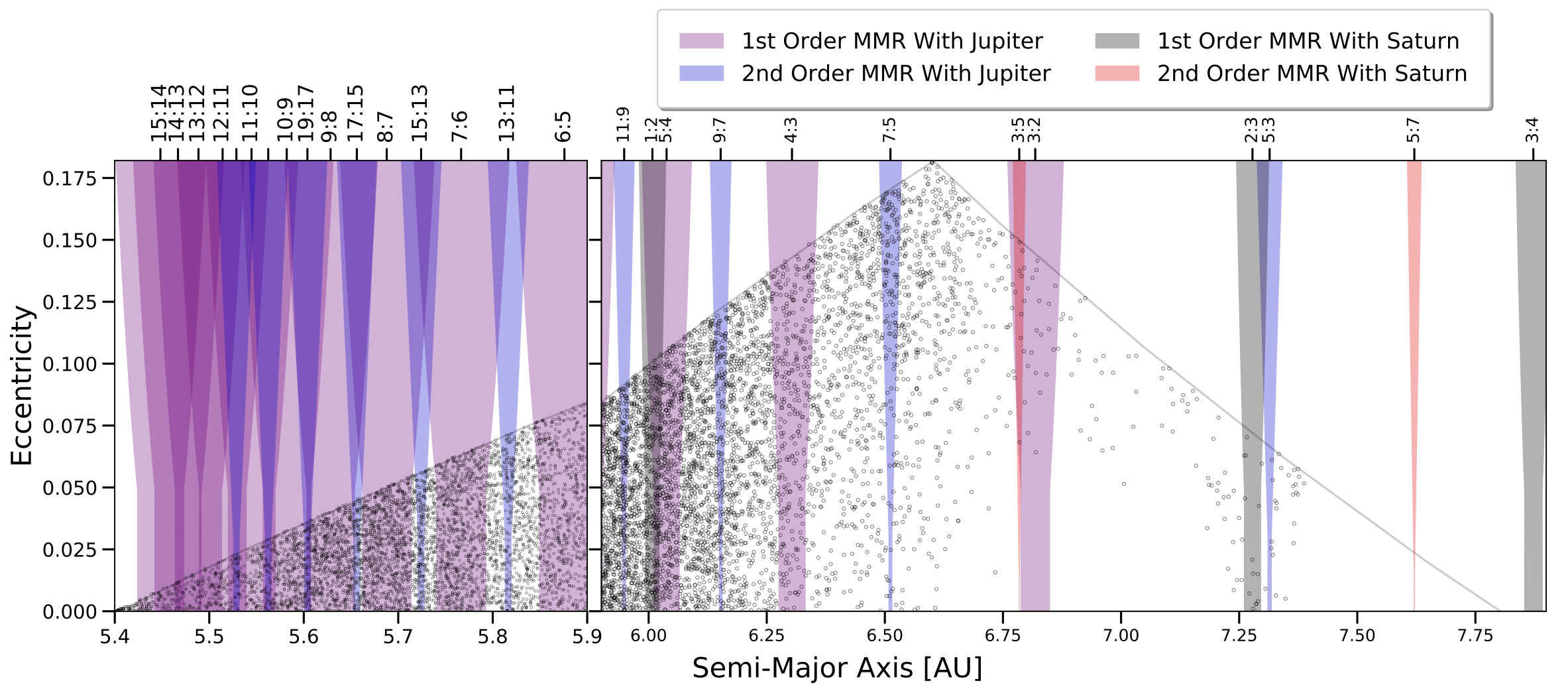} % this command will be ignored
\caption{Same as Figure \ref{Fig:MMR_saturn_jupiter}, but with the initial conditions of test particles that reach the $q<4$au within the next 1000 years. }\label{Fig:gateway_MMR_MC}
\end{center}
\end{figure*}

The top panel of Figure \ref{Fig:orbits_examples} shows a Centaur that begins close to the $9:7$ MMR with Jupiter at $a\sim6.15$au. Its eccentricity increases quickly, presumably due to the MMR, until it is no longer in the Gateway. The eccentricity subsequently lowers due to interactions with Jupiter, until is is dynamically excited to an eccentricity of $e\sim 0.3$ after about $\sim200$ years, and enters the region  $q<4$au. After this, the object is scattered back into the Centaur region, and eventually re-enters the Gateway. It is important to note that objects such as this will undergo volatile depletion after close approach to the Sun due to cometary fading \citep{Brasser2015}, and may be destroyed before re-entering the Centaur population. 

The second panel shows a more typical Centaur evolution, which exits the Gateway, and spends a significant fraction of the next 1000 years between the Gateway region and  $q<4$au. This object begins  close to the second order $7:5$ MMR with Jupiter at $a\sim6.5$au. In about 800 years, it's eccentricity is excited to the point where it attains $q<4$au. The third panel shows an extreme example of an object that begins with very low eccentricity and close to the inner edge of the  Gateway. Due to the overlapping MMRs in the region of low eccentricity where $a\in(5.4,5.65)$au, this object is immediately scattered to high eccentricity, and evolves within $\lesssim$ 500 years to a semi-major axis of $a\sim 11$au, before being scattered back into the inner Solar System and reaching $q<4$au after $\sim 900$ years. The fourth panel shows a Centaur that starts in the same dense region close to Jupiter but with higher eccentricity. This object is quickly scattered into the inner Solar System and reaches $q<4$au in less than 200 years for a short period, before getting scattered out to a higher eccentricity and semi-major axis. It is subsequently scattered back into the region where $q<4$au for a second time with   high eccentricity of $e\sim0.35$ after $\sim 400$ years, before being scattered back into the Centaur population. It is important to note that we simply infer resonant interactions, without verifying libration of resonant angles.

In Figure \ref{Fig:lifetimes_Gateway}, we show the initial conditions of objects in our numerical simulation that attain $q<4$ au within 1000 years.  In Figure \ref{Fig:gateway_MMR_MC}, we show these same objects in conjunction with the position and structure of the MMRs shown in Figure \ref{Fig:MMR_saturn_jupiter}. It is evident from these two figures that the region where $a\lesssim6$ au is the most densely populated, suggesting that the  MMRs are primarily responsible for injecting objects into the inner Solar System from the Gateway. The region where $a<5.7$au has a higher average time before reaching $q<4$au, as can be seen in  Figure \ref{Fig:lifetimes_Gateway}. However, this artifact is due to the large number of transitioning objects that  begin there. Due to the chaotic nature of the region close to Jupiter, the temporal evolution of these objects is very sensitive to their initial conditions. Since we have integrated over all of the orbital elements, we have densely sampled these regions in orbital element space. At larger semi-major axes, the population of Centaurs that transition is less densely populated. It is evident that the $4:3$ and $7:5$ MMRs with Jupiter are also efficient at injecting Centaurs into  region $q<4$au without requiring a close Jovian encounter. Of particular interest, the region where the  $3:2$ MMR with Jupiter and the  $3:5$ MMR with Saturn overlaps also produces objects that reach $q<4$au, as well as the region where the $5:3$ and $2:3$ MMRs are close to each other. %it is likely  conclude that the majority of Centaurs that are injected into the inner Solar System from the Gateway are from the inner regions of the Gateway closest to Jupiter. 

\begin{figure}
\begin{center}
\includegraphics[scale=0.4,angle=0]{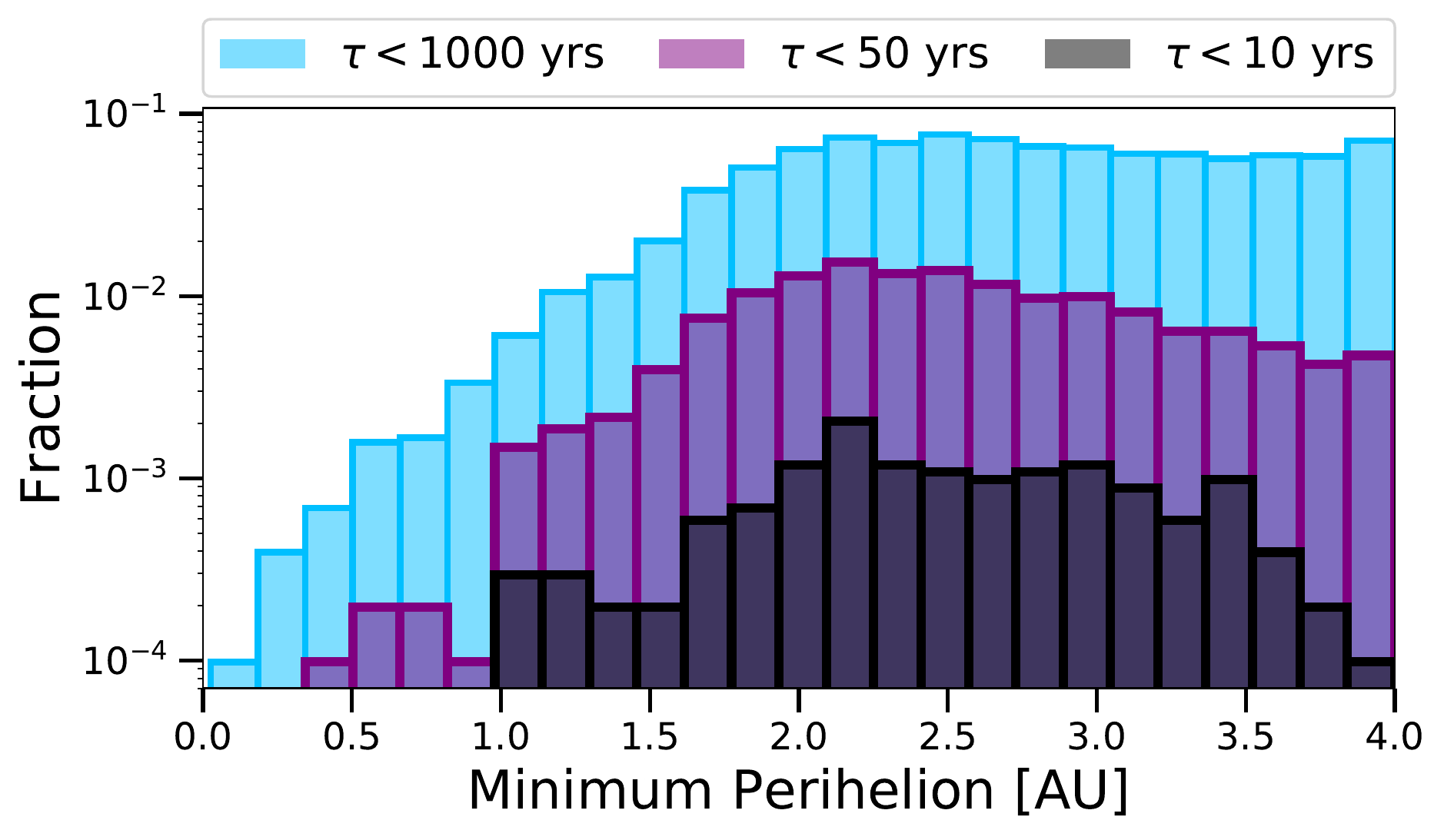}
\includegraphics[scale=0.4,angle=0]{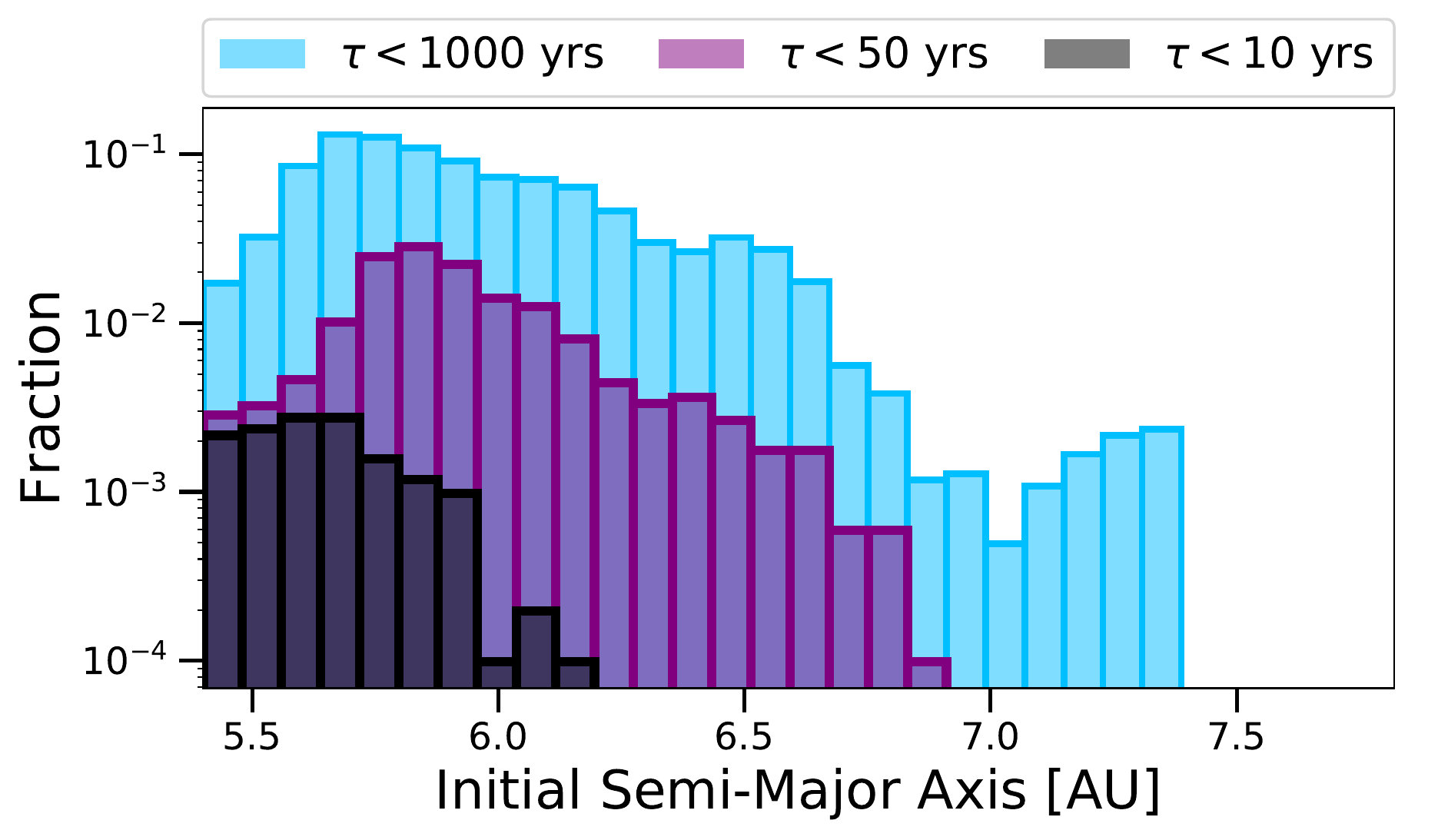}
\includegraphics[scale=0.4,angle=0]{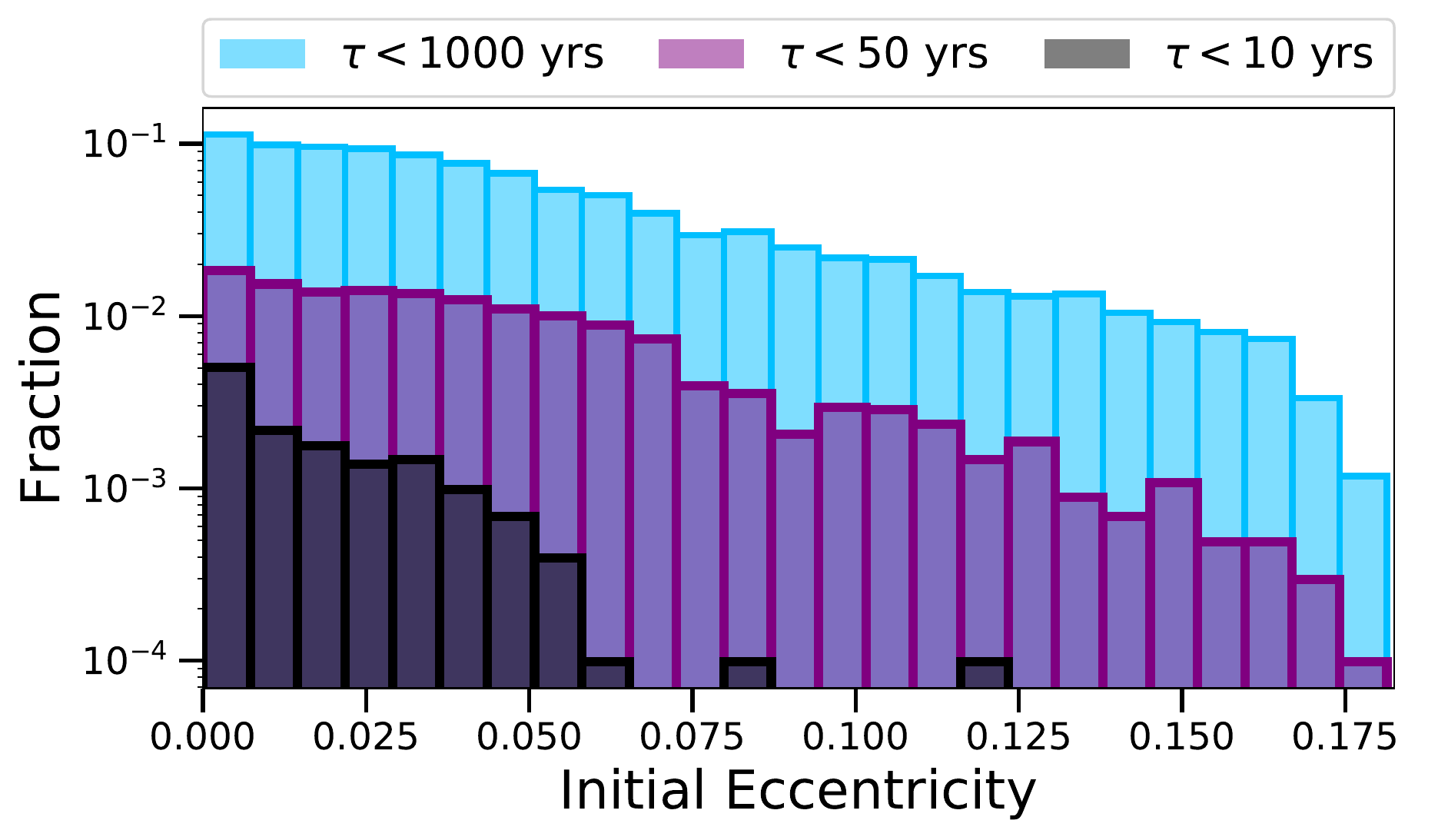}
\includegraphics[scale=0.4,angle=0]{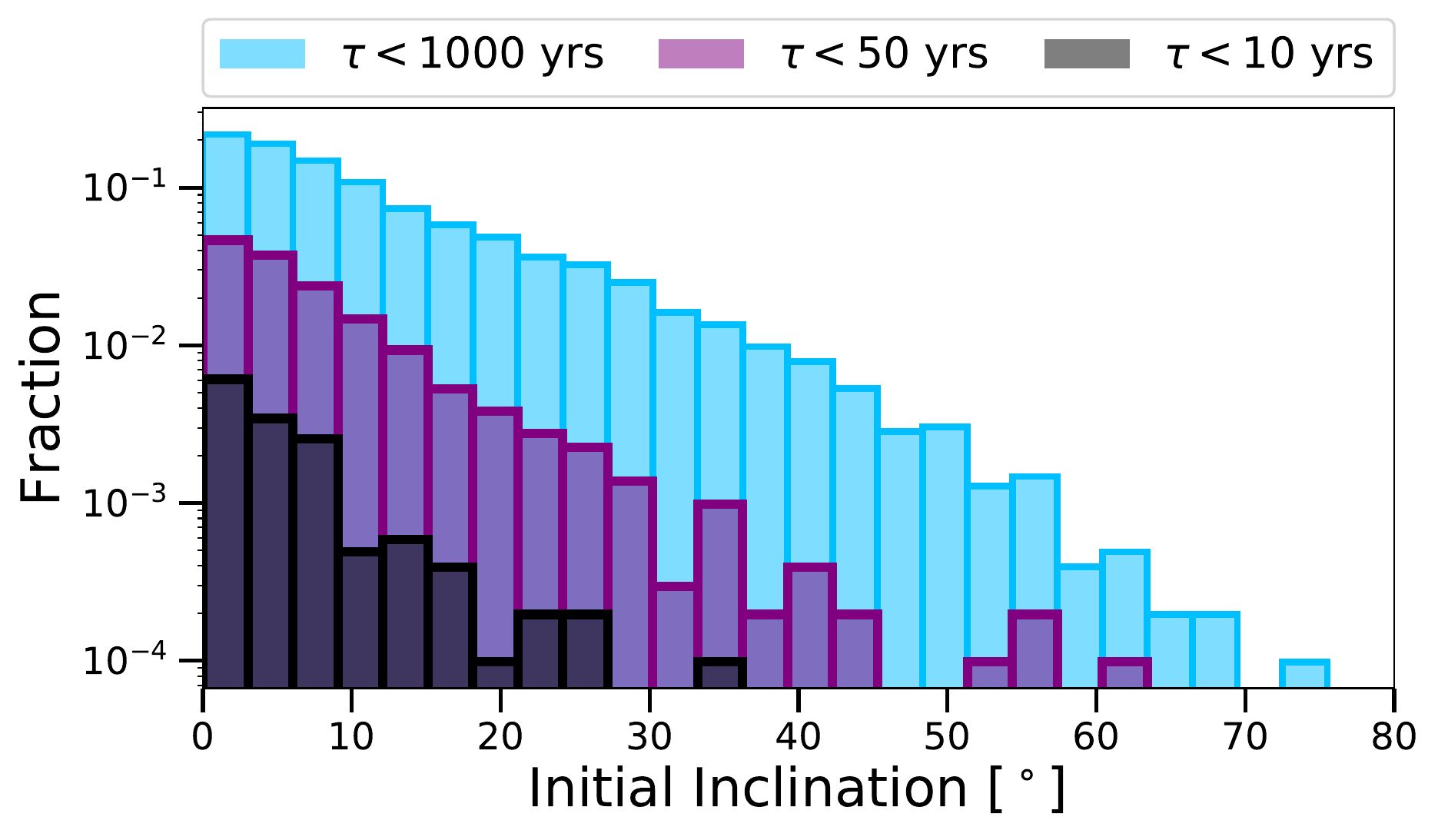}
\caption{The minimum perihelia value attained and initial semi-major axes, eccentricities and inclinations of objects that originate in the Gateway that reach $q<4$au. The  colored histograms represent the distributions of objects that reach $q<4$au within 1000, 50 and 10 years after 2021.}\label{Fig:histogram_lifetimes_gateway}
\end{center}
\end{figure}

In Figure \ref{Fig:histogram_lifetimes_gateway} we show histograms of the initial orbital elements and minimum perihelia attained for every object that reaches $q<4$au. It is important to note that the initial conditions do not imply that the test particles are formed at these locations. The purpose of these simulations is simply to identify regions of orbital element space that are amenable to quick transfer into the inner Solar System.  The majority of the simulated objects do not reach closer than 1au, but  $\sim 10^{-3}$ of the objects that reach $q<4$au  are scattered interior to Earth's orbit.  The majority of objects $(\sim 90\%)$ that transition start in the inner region of the Gateway close to Jupiter, with $a<6.5$ au, which provides a more  quantitative  representation of this feature than what is shown in Figure \ref{Fig:gateway_MMR_MC}. However, there are are  significant differences in the fraction of objects that reach $q<4$au  within the population that begins with $a>6.5$au. There are peaks in the histograms  close to $6.8$au and $7.25$au, the locations of MMRs, and there are no objects that start past $\sim7.3$au. This provides  numerical evidence that the MMRs are dynamically important for driving objects out of the Gateway region into the inner Solar System, corroborating the analysis presented in \citet{Bailey2009} and \citet{Tiscareno2003}. 

Objects that transition within the next 10 years are predominantly sampled from  orbits with eccentricity less than $e<0.05$. However, for objects that transition within 50 and 1000 years, the eccentricity distribution  samples the entirety of the Gateway region.  

Since the injection of objects into the inner Solar System appears to be driven by MMRs and direct scattering events, it is not expected that the initial inclination of the orbit should have an  effect on the population that transitions, since specifically these MMRs are not affected by inclination (although there exist many types of resonances for which inclination is important). This is consistent with the bottom panel of Figure \ref{Fig:histogram_lifetimes_gateway}, where the inclination distribution mirrors the initial conditions of our simulation. However, it  appears that  objects that transition in the next 10 years are primarily sampled from low inclination orbits with initial inclination $i<20^\circ$.

\subsection{Estimated Transfer Rates}\label{subsec:transferrates}
To translate the results of our numerical simulations into expected number of transitioning objects, we must estimate how many objects similar to LD2 currently occupy the Gateway. As noted in the previous subsesction, objects that experience close Solar encounters will undergo volatile depletion and potential destruction  due to cometary fading \citep{Brasser2015}, before re-entering the Centaur population. \citet{Sarid2019}  estimated that about 300 (1000) Centaurs with radii $R\gtrsim 1$km  currently reside in the  Gateway region for distributions assuming fading (and no fading).  LD2 itself has an absolute magnitude $H=12.2\pm0.8$ (JPL Horizons).  If this value represents the absolute magnitude of the nucleus, \citet{Steckloff2020} estimated that with albedos varying from 0.05 to 0.11, the nucleus of LD2 has a radius between $R\sim7-11$ km. \citet{Kareta2021PSJ} however, showed that the nucleus was likely to have a  radius of  $R\lesssim 1.2$km and $R\lesssim 0.8$km with the same albedos, using precovery DECam images. 

We assume that the Centaur distribution follows the power law used in \citet{Sarid2019} and \citet{Steckloff2020}, 
\begin{equation}\label{eq:sfd}
    dN_r=-k\alpha r^{-(\alpha+1)}dr\,,
\end{equation}
where $\alpha=3$ and $k=6.5\times10^6 \,\rm{km}^{-1}$. These authors estimated that there are currently $6.5\times10^6$  Centaurs with $R\gtrsim1$km (see Table 1 in \citet{Steckloff2020} and our Table \ref{tab:definitions}), and estimated that the Gateway  currently had $\sim 240$ objects with radii $R\gtrsim 1$km. However, if the nucleus of LD2 is smaller than previously noted, and $R_{LD2}\sim0.8$ km, while reaching the apparent magnitude it did, then it is possible that it is representative of a larger population than previously estimated \citep{Steckloff2020}. Integrating Equation \ref{eq:sfd} from 0.8 km increases the number of Centaurs with $R\gtrsim 0.8$km to be of order $2\times10^7$, which is a factor of $\sim$ 3 greater than the estimates in \citet{Steckloff2020}, after accounting for fading as in \citet{Brasser2015}. With this assumption, the population of LD2 like objects in the Gateway could be $\sim700$, although this number is very uncertain. 

We use the fiducial estimates of $\sim240$ and $\sim700$ LD2-like objects to normalize our distributions of objects that reach low perihelia in our simulation to estimate the number of objects that will transition out of the Gateway region and reach $q<4$au. Table \ref{tab:gateway_transition_numbers} shows the numbers of objects that will transition soon, with these two assumed populations of objects in the Gateway. We show the number of objects produced with both estimates, and for objects that reach $q<4$ and $q<3.5$ au at some point in the next 1000 years. %In the next 25 years, these estimates imply that at least 1 centaur will transition out of the Gateway and reach a perihelion distance of $q<3.5AU$, like LD2. If LD2's nucleus is on the smaller end, then we can expect one such object transitioning in the next 10 years, and roughly 5 such objects transitioning in the next 25 years.  

The estimates of the number of transitioning objects  presented here reflect the uniform initial conditions.  There is currently no observational evidence that supports this hypothesis. Moreover, given the chaotic nature of the Gateway region, it is likely that the regions that are amenable to orbital transfer are  severely depleted. This should  revise-down the inferred rates that objects will be scattered into the inner Solar System. In order to account for this depletion in our estimates, we apply the  cuts described below to revise-down our estimates.

\citet{Roberts2021} investigated the transfer of objects out of the region between Jupiter and Saturn.  They defined this as the near Centaur region, which is larger than the  Gateway (see Table \ref{tab:definitions}). Although they used different assumptions from \citet{Sarid2019} and investigated a larger parameter space, the  population estimates were in general agreement in both studies for most cases.  \citet{Roberts2021} normalized their populations by matching the numerical injection rates of OJFCs to the observed population (see the discussion in Subsection \ref{subsec:definitions}). \citet{Rickman2017} estimated that the injection rate was $8.4\pm1.7\times 10^{-3}$ yr$^{-1}$, which was in good agreement with the rates presented in Table 2 of \citet{Roberts2021}, $14.6\pm3.9\times 10^{-3}$ yr$^{-1}$, $13.3\pm2.7\times 10^{-3}$ yr$^{-1}$ and $9.6\pm4.9\times 10^{-3}$ yr$^{-1}$ for three different source populations. 

Our most comparable rate to these published values is $\sim .11$ yr$^{-1}$, which represents the rate of objects that attain $q<4$au, if LD2's radius is $R\gtrsim 1$km. Of the objects that reach $q<4$au, only $\sim1/4$ of these satisfy the stricter perihelia criteria for the definition of OJFCs. Therefore, our estimated rates are likely overestimates by a factor of $f_{dep}\sim 3$. We attribute this overestimation to the fact that the regions in the Gateway that are amenable to quick transfer should be depleted. Moreover, Table 1 in \citet{Sarid2019} demonstrates that for the entire Centaur population that they simulated, $\sim50\%$ of the objects that became JFCs with $q<3$au had phases in the Gateway before transitioning. To match the OJFC injection rate, we therefore revise-down the number estimates by an additional factor of $\sim2$. Therefore, in the final two rows of Table \ref{tab:gateway_transition_numbers}, we show the numbers that are revised down by dividing by  $f_{dep}\sim6$. With these revised estimates, it is still possible that 1-2 objects will transition in the next 25-50 years. It is important to note that \citet{Steckloff2020} found that the median frequency with which objects transition from the Gateway to the JFC population was once every $\sim2.7$ years and $\sim73$ years, if the radius of LD2 is $R\gtrsim 1$km and $R\gtrsim 3$km, respectively. These estimates were calculated using detailed numerical simulations that tracked a large number of TNO test particles through the Centaur region presented in \citet{Sarid2019}. Therefore, it is plausible that there will be more than 1-2 objects that transition in the next 25-50 years. If such an object is detected in the future, it would be an intriguing target for a rendezvous mission such as the one described in \S \ref{sec:missions}.

\begin{table}[]
    \centering
    \begin{tabular}{c|c|c|c|c|c|c}
          Years & 1000 &100& 50 &25& 10\\
        \hline
         $R_{LD2}\sim 0.8$km $q<3.5$au& $ 92$ & $ 20$ &$10$& 
        $5$& $1$&\\
         $R_{LD2}\gtrsim 1$km $q<3.5$au& $30$ & $7$ &$ 3$& 
        $1.5$& $0.5$&\\
        $R_{LD2}\sim 0.8$km $q<4$au& $114$ & $34$ &$17$& 
        $ 8$& $2$&\\
         $R_{LD2}\gtrsim 1$km $q<4$au& $36$ & $11$ &$6$& 
        $3$& $0.5$&\\
        $R_{LD2}\sim 0.8$km $q<4$au $/f_{dep}$ & $19$ & $5$ &$2.5$& 
        $ 1$& $.15$&\\
         $R_{LD2}\gtrsim 1$km $q<4$au $/f_{dep}$& $6$ & $1.5$ &$1$& 
        $0.5$& $0.08$&\\
    \end{tabular}
    \caption{The predicted number of Gateway objects that reach $q<4,3.5$au within the next 1000, 100 50, 25 and 10 years.}
    \label{tab:gateway_transition_numbers}
\end{table}

\section{Centaur Population Simulation}\label{sec:fullcentaur}

\begin{figure*}
\begin{center}
\includegraphics[scale=0.5,angle=0]{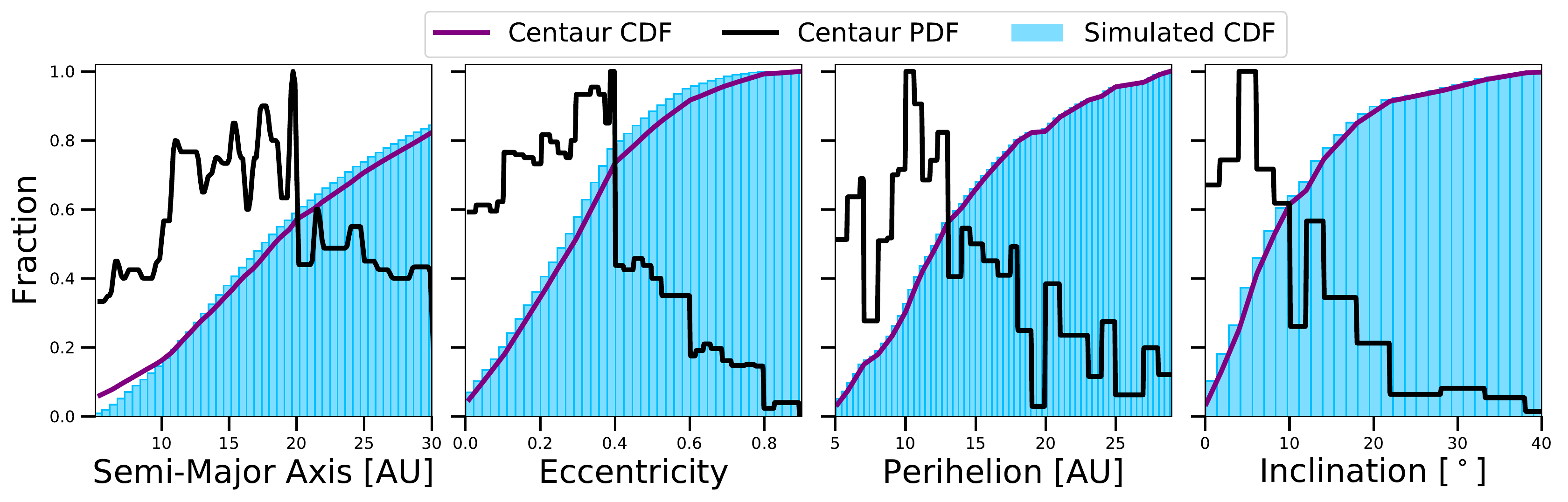}
\caption{Initial conditions of our dynamical simulations. The purple lines show the CDFs  of orbital elements for Centaurs from Figure 13 in \citet{DiSisto2007}, and the associated PDFs are shown in black lines.  The blue histograms are the initial conditions of our simulations.} \label{Fig:IC_disisto}
\end{center}
\end{figure*}

\begin{figure}
\begin{center}
\includegraphics[scale=0.35,angle=0]{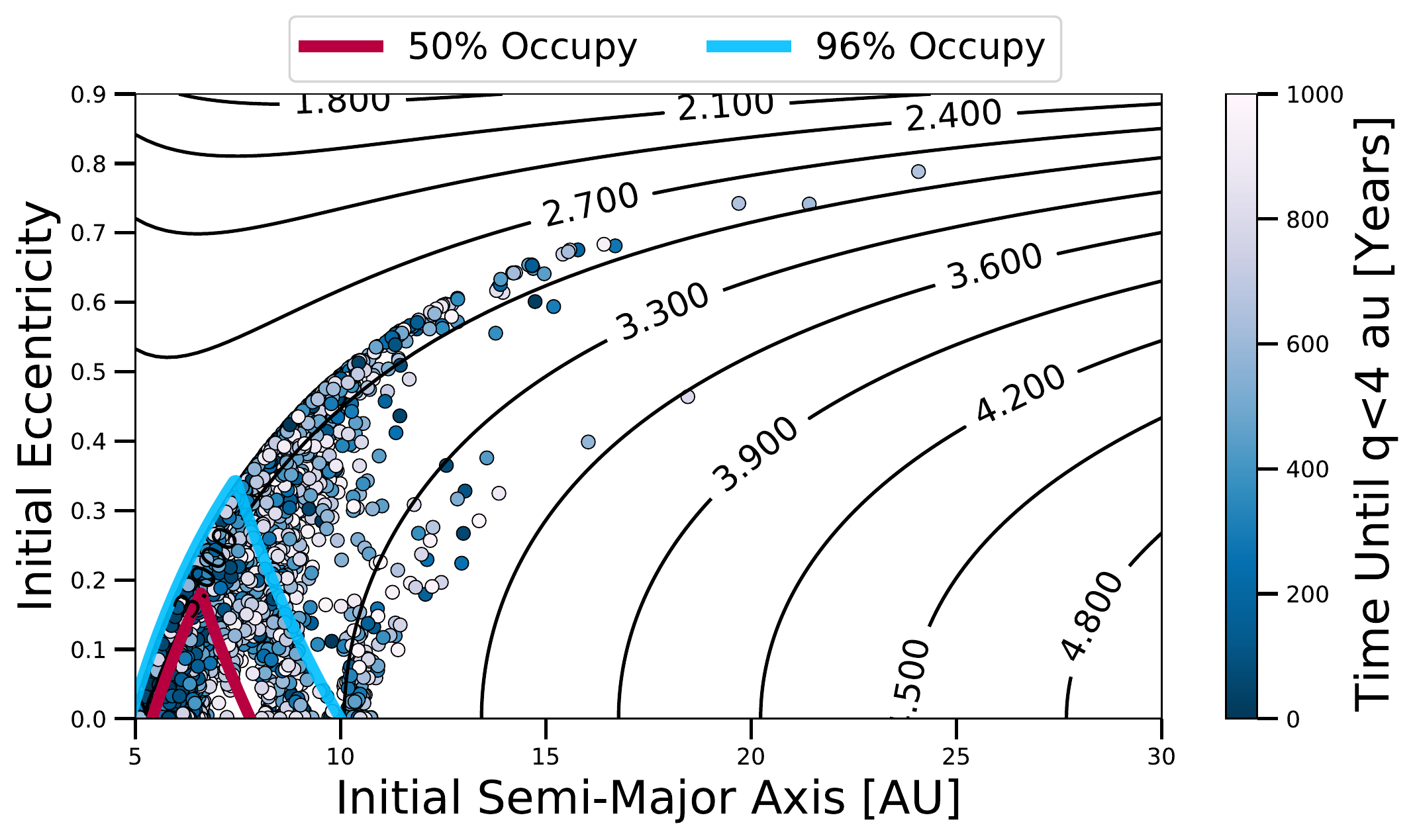} % this command will be ignored
\caption{Initial conditions of  test particles in our simulation that reach  $q<4$au within 1000 years.  The color of each point indicates the time elapsed before $q<4$au. Contours of the Tisserand parameter with respect to Jupiter are overplotted. The Gateway and the region defined by $q>a_J-R_{HJ}$ and $Q<a_S+R_{HS}$ are plotted in red and blue solid lines, where $a_S$ is Saturn's semi-major axis and $R_{HJ}$ and $R_{HS}$ are Jupiter and Saturn's Hill radii. Of the  particles that reach $q<4$au, $50\%$ and $96\%$ occupied these regions prior to their journey into the inner Solar System.  }\label{Fig:liftimes_fullpopulation}
\end{center}
\end{figure}

\begin{figure}
\begin{center}
\includegraphics[scale=0.45,angle=0]{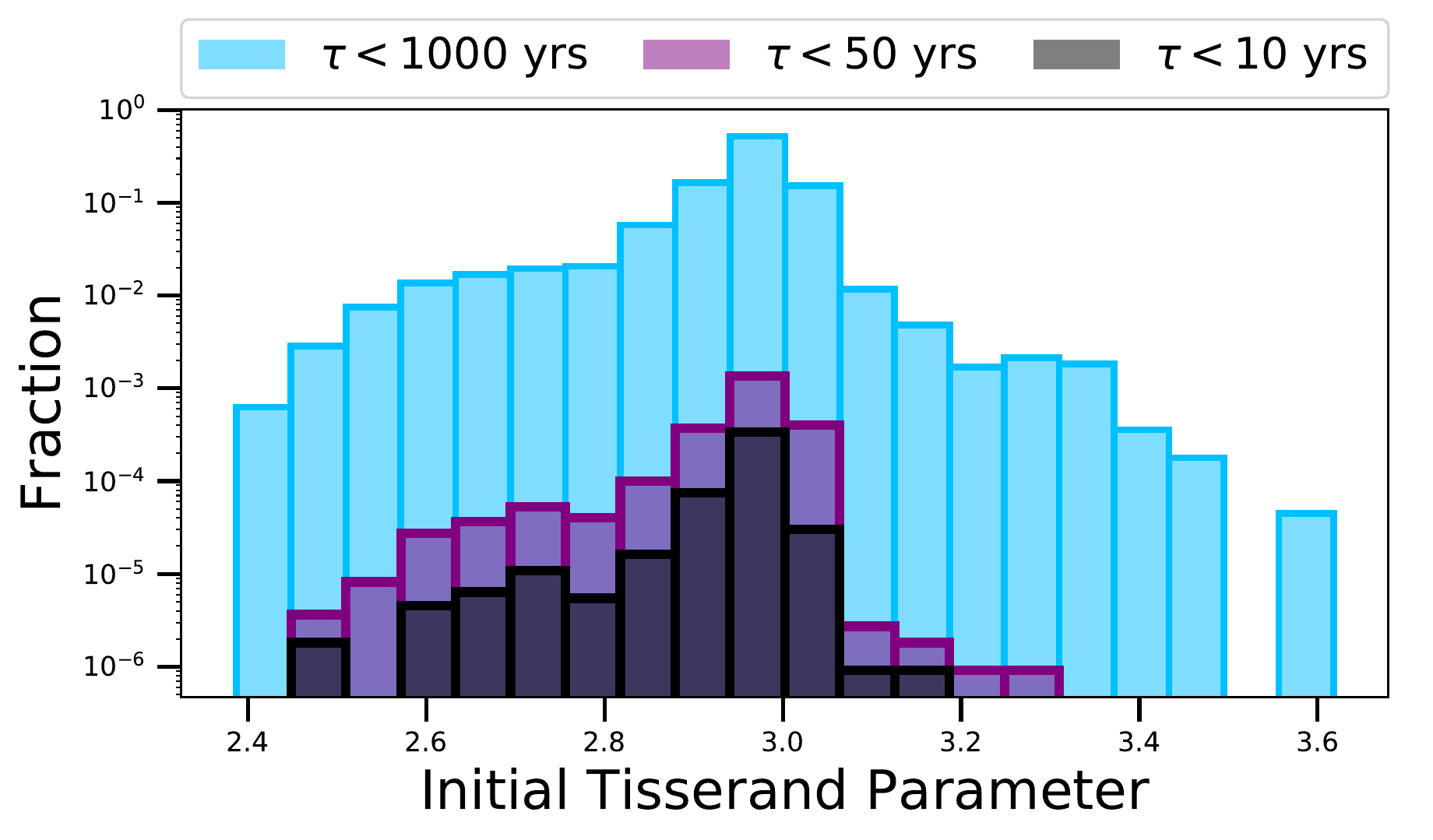} % this command will be ignored
\caption{The distribution of the initial Tisserand parameter with respect to Jupiter for objects that  reach $q<4$au within 1000 years.}\label{Fig:tisserand_histogram}
\end{center}
\end{figure}

\begin{figure}
\begin{center}
\includegraphics[scale=0.4,angle=0]{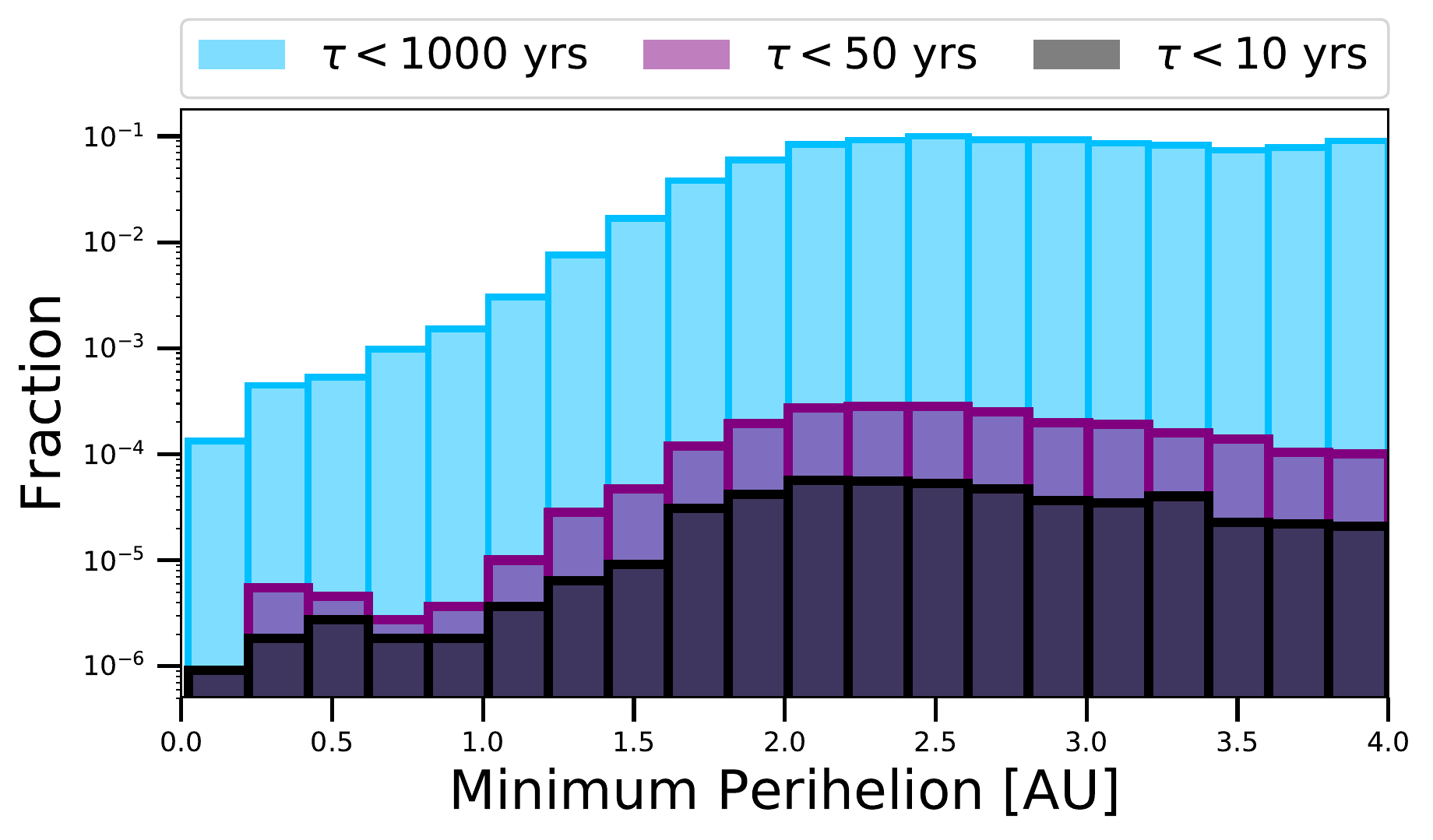}
\includegraphics[scale=0.4,angle=0]{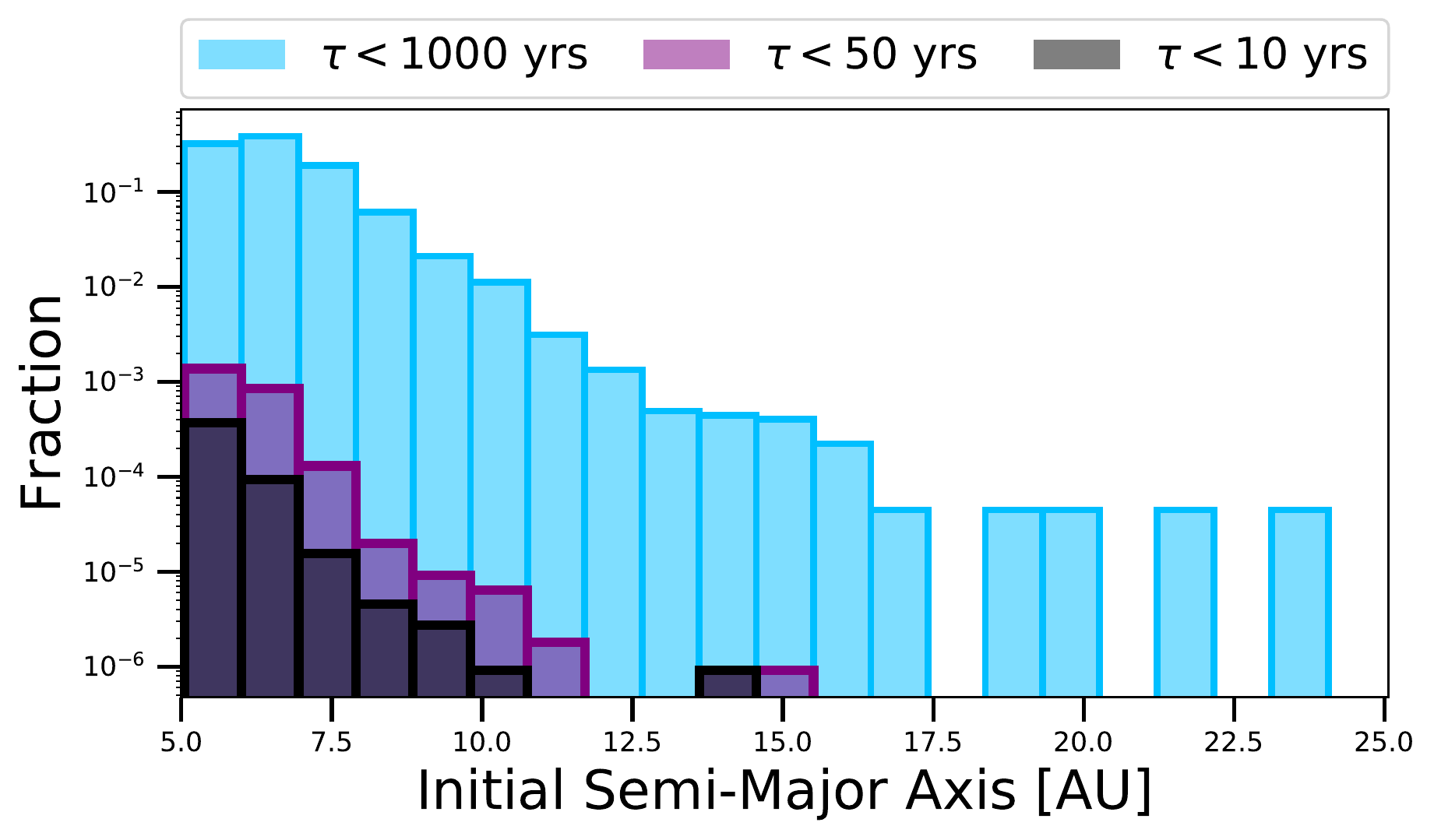}
\includegraphics[scale=0.4,angle=0]{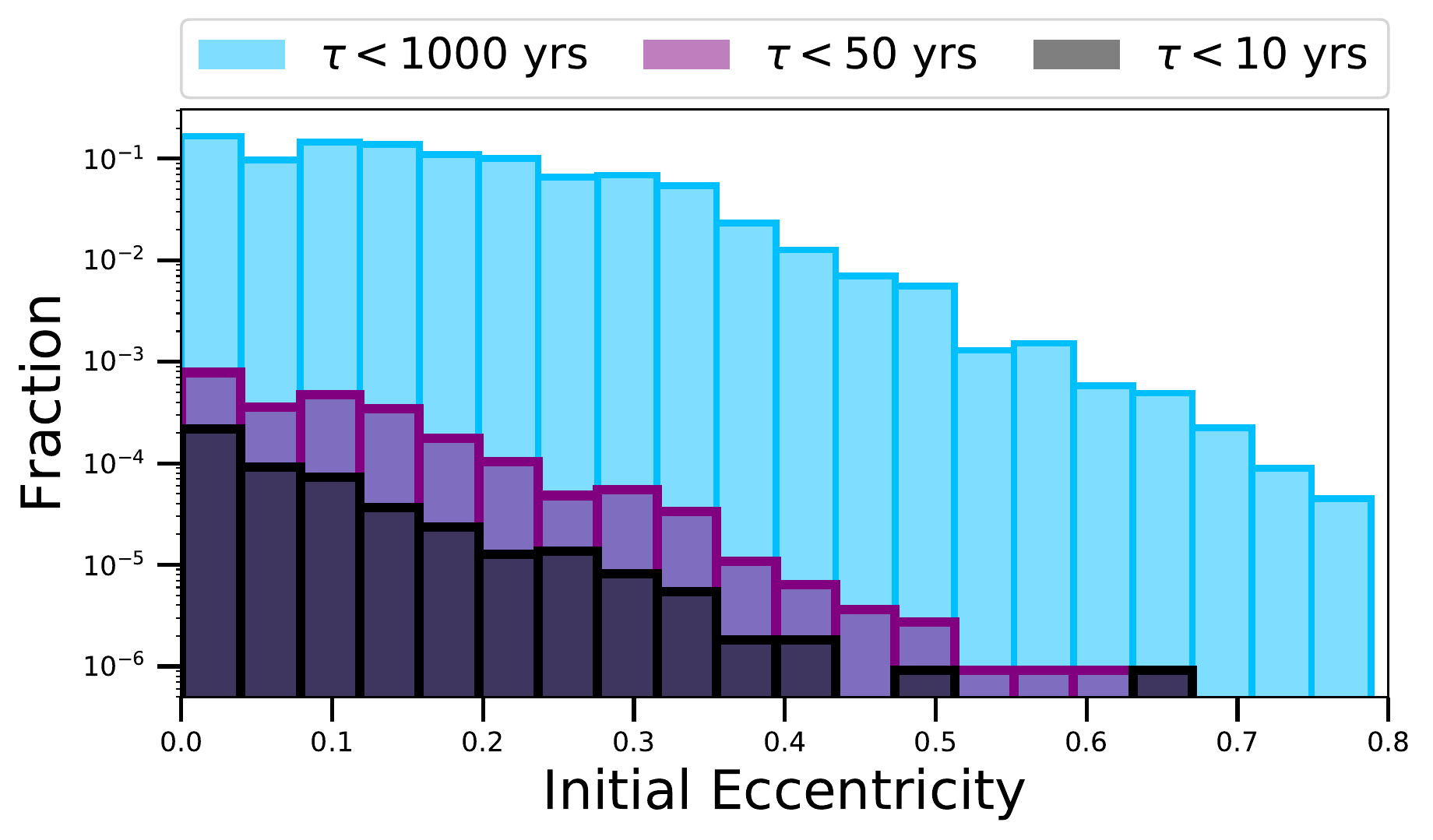}
\includegraphics[scale=0.4,angle=0]{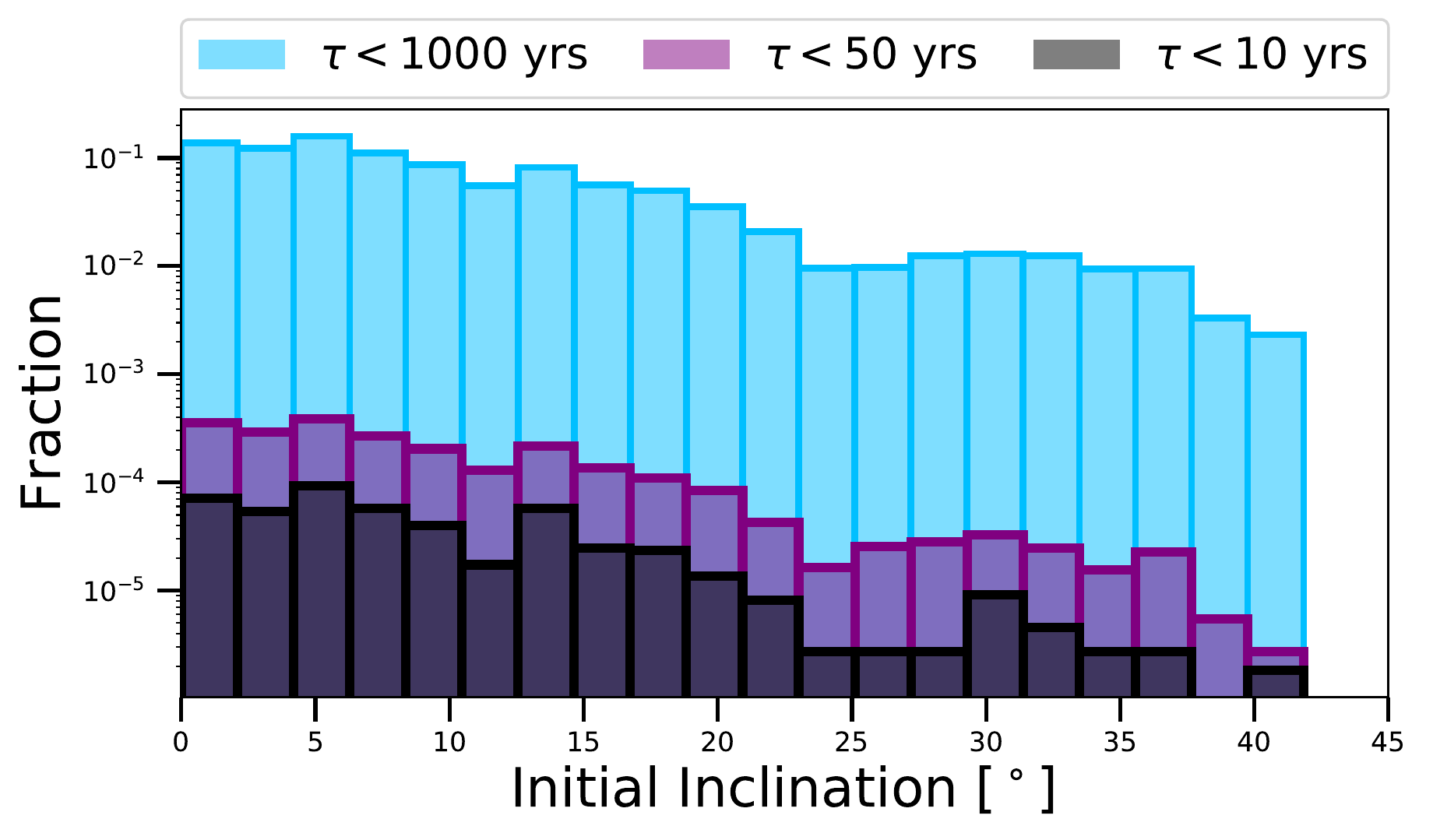}
\caption{The  distributions of minimum perihelion distance attained and initial orbital elements for all of the simulated test particles that reach $q<4$au, as in Figure \ref{Fig:histogram_lifetimes_gateway}.  The different colors show test particles that reach $q<4$au within 1000, 50 and 10 years from 2021.}\label{Fig:histogram_lifetimes_fullpopulation}
\end{center}
\end{figure}

\begin{figure}
\begin{center}
\includegraphics[scale=0.45,angle=0]{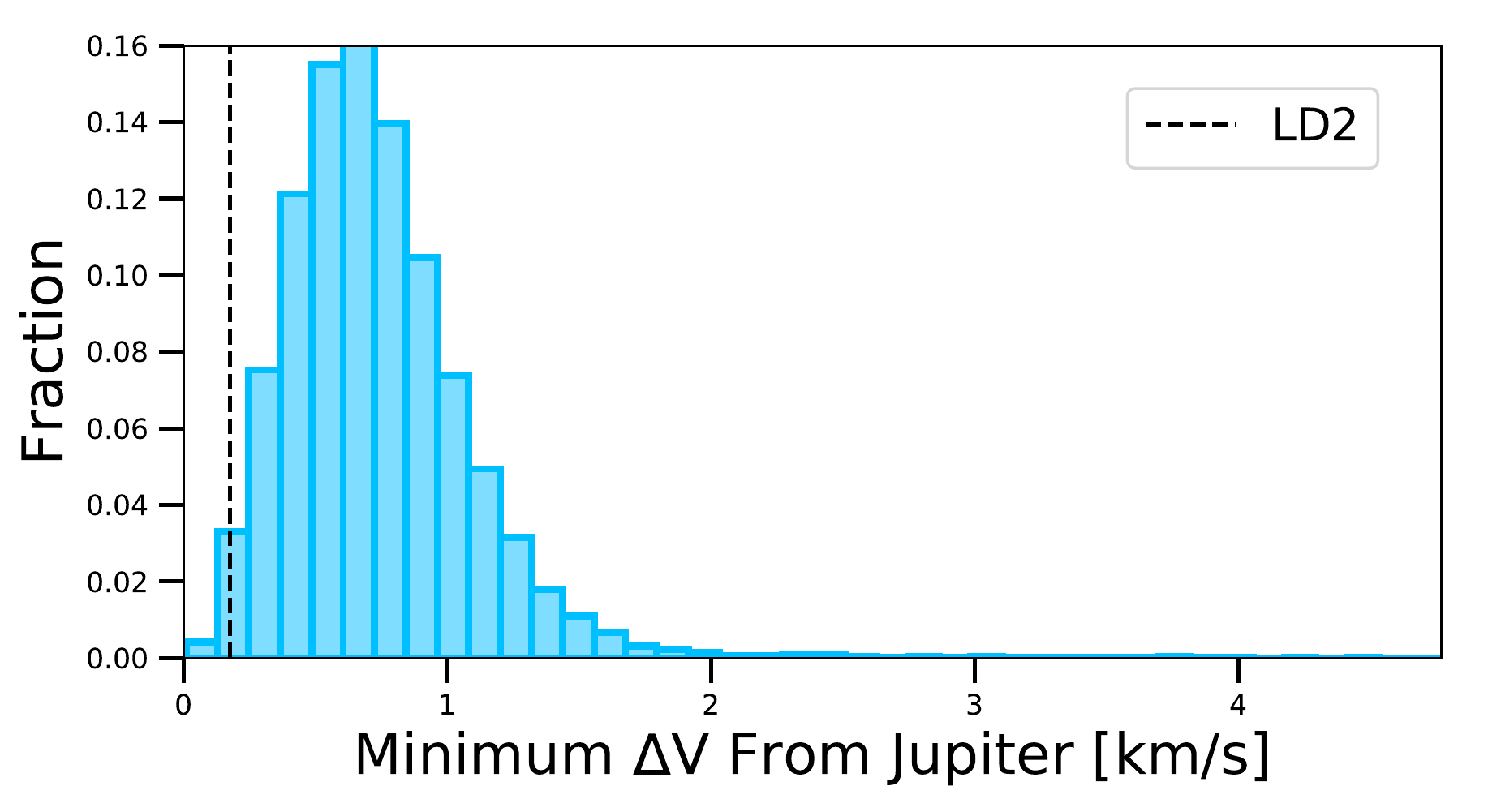}
\caption{The distribution of the minimum $\Delta $V with respect to Jupiter for simulated test particles that reach $q<4$au}\label{Fig:hypthothetical_mission}
\end{center}
\end{figure}

As we shall demonstrate in \S \ref{sec:missions}, it will be feasible to perform a rendezvous mission with LD2. In \S \ref{sec:simulations}, we showed that it is possible that of order $\sim 1$ additional target may transition into the inner Solar System in the next $\sim50$ years, with large uncertainties. \citet{Sarid2019} demonstrated that $\sim50\%$ of objects that attain $q<3$au did \textit{not} have a Gateway phase prior to transitioning. In order to contextualize LD2 within the entire Centaur population, and identify the dynamics of potential targets that do not occupy the Gateway prior to reaching $q<4$au, we perform N-body simulations of a larger population of Centaurs in this section. 

The distribution of orbital elements for Centaurs is not well constrained by observations and has  been probed via theoretical N-body calculations \citep{DiSisto2007,Disisto2020,Roberts2021}. Because we are primarily interested in the dynamics of objects as they reach the inner Solar System, we draw the initial conditions for test particles from the steady state Centaur population simulated in \citet{DiSisto2007}. Although the initial conditions in this section are more physically motivated than in the previous section, the fractional area of orbital element space that we simulate here is much larger.  This  prevents us from commenting on population-level statistics in this section. It would be worthwhile to compare these simulations to those that tracked the evolution of test particles from the Kuiper belt, but this is outside the scope of this paper. 

\begin{figure*}
\begin{center}
\includegraphics[scale=0.9,angle=0]{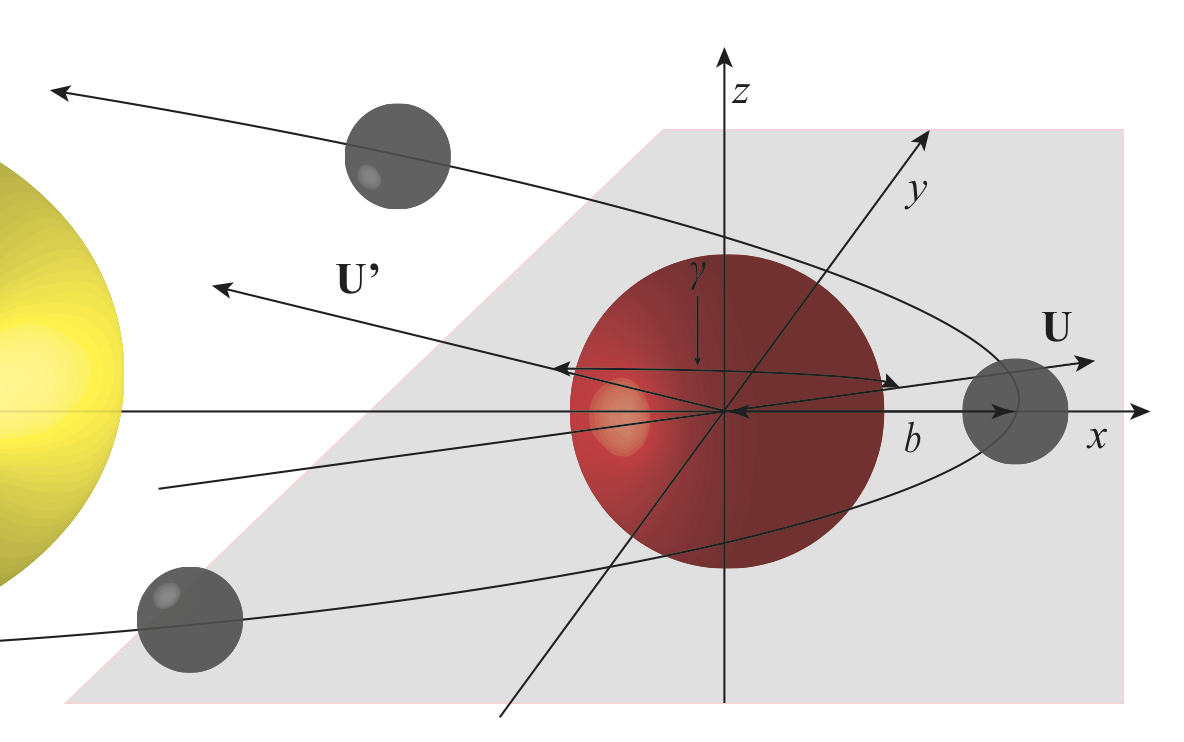}
\caption{The geometry of a scattering event of an object with Jupiter. The Sun, Jupiter and test particle are shown in yellow, red and grey respectively, and are not drawn to scale. The test particle  is show when it is (1) incoming, (2) at closest approach to Jupiter and (3) outgoing. The incoming and outgoing velocity vectors ${\bf U}$ and ${\bf U'}$, impact parameter, $b$, and deflection angle, $\gamma$, are indicated.}\label{Fig:schematic_scatter}
\end{center}
\end{figure*}

We generate an underlying population of $\sim 1.2$ million  test particles, and then perform N-body integrations as in \S\ref{sec:simulations}. In order to generate the initial orbital elements, we used the cumulative distribution functions (CDFs) of the orbital elements presented in Figure 13 in \citet{DiSisto2007}, digitized  using \textit{Automeris} \citep{ank17}. These CDFs represented the steady state population of Centaurs integrated using N-Body simulations that begun with the observed and theoretical population of SDOs. In Figure \ref{Fig:IC_disisto}, we show the CDFs and resulting PDFs of the semi-major axis, eccentricity, perihelion and inclination of this steady state population. 

In order to generate initial conditions for the test particles, we follow an iterative procedure in which we (1) draw $q$ from the CDF, (2) draw $e$ from the CDF, and  (3) calculate the resulting $a$. If $a<5$au or $a>50$au, we redo steps 2 and 3 until we find an orbit that has $a\in (5,50)$au. It is important to note that this procedure does not generate a population of objects that strictly follows the definition of a Centaur in \citet{Jewitt2009}, or captures the interdependencies in the underlying CDFs.  However, we are primarily interested in identifying the initial conditions of objects that reach $q<4$au in this entire region, despite the various literature definitions. We draw $i$ from the CDF in \citet{DiSisto2007}.  The blue histograms in Figure \ref{Fig:IC_disisto} show our resulting initial conditions, which provide a reasonable approximation to the theoretical CDFs for the purposes of our dynamical investigation. We randomly draw longitude of periapse, $\omega$ and longitude of ascending node, $\Omega$. We integrate each test particle for 1000 years starting from 2021, along with all of the terrestrial and giant planets.  

In Figure \ref{Fig:liftimes_fullpopulation} we show the distribution of test particles that transition to $q<4$au. We plot  contours of the Tisserand parameter with respect to Jupiter, $T_J$ (excluding the inclination dependence). Values of $T_J\sim3$ are a reasonable indication that an object  will be injected  into the inner Solar System over the course of the simulation.  Although the simulations included objects with orbits out to that of Neptune, the majority of objects that reach the inner Solar System in 1000 years begin within the orbit of Saturn. 

We verified that the structure of the subset of initial conditions that led to test particles reaching $q<4$au within the Gateway region was consistent with those presented in Figure \ref{Fig:lifetimes_Gateway}. Moreover, we found that of the particles that reach $q<4,3.5$ and $3$au,  $46.8,49.8$ and $50.5\%$ had Gateway phases prior to reaching $q<4$au, respectively. This is in good agreement with the numbers presented in Table 1 of \citet{Sarid2019}. If we revise the definition  of the Gateway, and instead consider the region where $q>a_J-R_{HJ}$ and $Q<a_S+R_{HS}$, where $a_S$ is Saturn's semi-major axis and $R_{HJ}$ and $R_{HS}$ are Jupiter and Saturn's Hill radii, of the particles that reach $q<4,3.5$ and $3$au, $95.8,97.1$ and $97.5\%$ occupied this region prior to reaching $q<4$au,  respectively. The Gateway and this new region are indicated in Figure \ref{Fig:liftimes_fullpopulation}.

In Figure \ref{Fig:tisserand_histogram}, we show the distribution of the initial Tisserand parameter with respect to Jupiter for every object that reaches $q<4$au. As opposed to in Figure \ref{Fig:liftimes_fullpopulation}, the Tisserand parameter is  calculated with the inclination dependence of the initial test particle. The majority of objects ($>90\%$) that reach $q<4$au in the next 1000 years begin with $T_J\in(2.8,3.1)$, and all objects have $T_J\in(2.3,3.7)$. While almost all objects that reach the inner Solar System in the next 50 years have $T_J\in(2.8,3.1)$, the distribution of objects that transition in the next 1000 years has asymmetric tails at larger and smaller $T_J$. 

In Figure \ref{Fig:histogram_lifetimes_fullpopulation} we show the distributions of the minimum perihelion reached and initial orbital elements for objects that reach $q<4$au, as in Figure \ref{Fig:histogram_lifetimes_gateway}. The minimum perihelion distribution is similar to that shown in Figure \ref{Fig:histogram_lifetimes_gateway}, indicating that the minimum $q$ is generally independent of the initial  distance. Objects that are initialized closer to Jupiter are more likely to be scattered into the inner Solar System, and the majority ($>90\%$) of test particles that reach $q<4$au in the next 1000 years start between the orbits of Jupiter and Saturn. However $\sim 10\% $  of objects that reach the inner Solar System in the next 1000 years  start with semi-major axis exterior to that  of Saturn. As can be seen in Figure \ref{Fig:liftimes_fullpopulation},  objects that begin with $a>11$au also have high eccentricity, $e\gtrsim0.2$, and therefore have lower perihelia values. The inclination distribution roughly matches the initial conditions shown in Figure \ref{Fig:IC_disisto}, so as in \S \ref{sec:simulations}, we conclude that the mechanisms that drive the injection into the inner Solar System is generally independent of inclination.

It is possible that additional targets for a rendezvous mission will transition into the inner Solar System before LD2 will be detected in the future. While the simulations presented in this section are not appropriate for calculating the transition rates, they are useful for approximating what the trajectories of these targets will be. As we shall demonstrate in \S \ref{sec:missions}, the low $\Delta {\rm V}$ with respect to Jupiter makes a Jupiter-Sun Lagrange point a promising loitering location for a rendezvous spacecraft. In Figure \ref{Fig:hypthothetical_mission}, we show the distribution of $\Delta$V with respect to Jupiter of all of the objects in our simulations that reach $q<4$au. Almost all ($\sim 98\%$) of our simulated test particles have $\Delta {\rm V}<2$ km/s with respect to Jupiter, making them feasible targets for an orbit matching rendezvous similar to the one that we propose for LD2 in \S \ref{sec:missions}. It is important to note that this distribution represents the entirety of the objects in our simulated population that reach the inner Solar System, including objects that do not have close encounters with Jupiter. It appears that the low $\Delta$V with respect to Jupiter is a  feature of all objects that reach $q<4$au, independent of their orbital history. The distribution of $\Delta $V is roughly log-normal, and has a median  at $\Delta {\rm V}=0.68$km/s.

\section{Changes in Orbital Elements due to Gravitational Scattering by Jupiter}\label{sec:scattering}

\subsection{General Formalism}

We calculate the changes in orbital elements of low inclination objects that are scatted by Jupiter, to  verify the numerical results presented in the previous section, and to provide a physical explanation for this mechanism of generating objects that reach $q<4$au with low $\Delta$V with respect to Jupiter. The methodology presented here can be applied to easily predict whether  objects detected in the future are likely to be scattered into the inner Solar System.

\citet{Carusi1990} presented an analytic method for determining the outgoing orbital elements of a trajectory following a close encounter with a massive perturber. This work drew heavily on the methods presented by \citet{Opik1951} and \citet{Opik1976}, which was validated by \citet{Greenberg1988}. These works extend the simpler case of the 2D trajectory perturbation from a gravitational encounter, as in Chapter 2 of \citet{Murray1999}, and subsequently studied extensively by \citet{Longcope2020} with application to the trajectory of the Parker Solar Probe \citep{Guo2010}.

Specifically, for an initial orbit characterized by $(a,e,i,\omega,\Omega)$, \citet{Carusi1990} presented an analytic method to determine the post-encounter orbital elements, denoted with a prime, $(a',e',i')$ for a given impact parameter, $b$ and deflection angle, $\gamma$ between the incoming velocity vector in the frame co-rotating with Jupiter, ${\bf U}$, and post-encounter velocity vector, ${\bf U}'$ . In this section, we will utilize their methodology to examine the orbits of objects that are scattered by Jupiter to identify the trajectories that lead to the generation of objects with  $q<4$au. %We reproduce some of the relevant equations presented in their paper here. 

The geometry of the close encounter is depicted in Figure \ref{Fig:schematic_scatter}. The pre-encounter orbital elements $(a,e,i)$, with the assumption that $\Omega=0$, isomorphically map to the three components of the incoming velocity vector $(U_x,U_y,U_z)$. The coordinates are centered on the perturber, and the x-axis is parallel to the vector between the Sun and the perturber. The y-axis is in the instantaneous direction of the motion of the perturber at closest approach, and the z-axis is in the direction of the perturber's angular momentum vector. These relationships are given in Equation 8 and 9 of \citet{Carusi1990}, and are,

\begin{equation}\label{eq:orb_to_U}
    \begin{dcases}
    U_x=\big(2-\frac{1}{a}-a(1-e^2)\big)^{1/2} \\
    U_y= \sqrt{a(1-e^2)}\cos(i)-1 \\
    U_z= \sqrt{a(1-e^2)}\sin(i) \\
    \end{dcases}     
    \, ,
\end{equation}
and
\begin{equation}\label{eq:U_to_orb}
    \begin{dcases}
    a=\frac{1}{1-|{\bf U}|^2-2U_y} \\
    e= \big(|{ \bf U}|^4+4U_y^2+U_x^2(1-|{\bf U}|^2-2U_y)+4|{\bf U}|\big)^{1/2} \\
    i= \sin^{-1}\big(U_z^2/(U_z^2+(1+U_y)^2)\big) \\
    \end{dcases}     
    \, .
\end{equation}

These equation rely on the angles $\theta$ and $\phi$, which determine the direction of ${\bf U}$, and may be calculated using Equations 4 and 5 in \citet{Carusi1990},
\begin{equation}\label{eq:theta}
   \cos\theta=\frac{1-|{\bf U}|^2-1/a}{2|{\bf U}|}\,,
\end{equation}
and 

\begin{equation}\label{eq:phi}
   \tan\phi=\frac{\pm1}{\sin i} \sqrt{\frac{2a-1}{a^2(1-e^2)}-1}\,\,.
\end{equation}

Given an impact parameter, $b$, ${\bf U}$ maps to ${\bf U'}$ via a rotation of the angle, $\gamma$, in the direction of the angle $\psi$ from the meridian containing ${\bf U}$. The angle $\gamma$ is defined in Equation 10 of \citet{Carusi1990},

\begin{equation}\label{eq:gamma}
   \tan(\gamma/2)=\bigg(\frac{M_J}{M_\odot }\bigg)\frac{1}{b|{\bf U}|^2}\,,
\end{equation}
\begin{figure}
\begin{center}
\includegraphics[scale=0.7,angle=0]{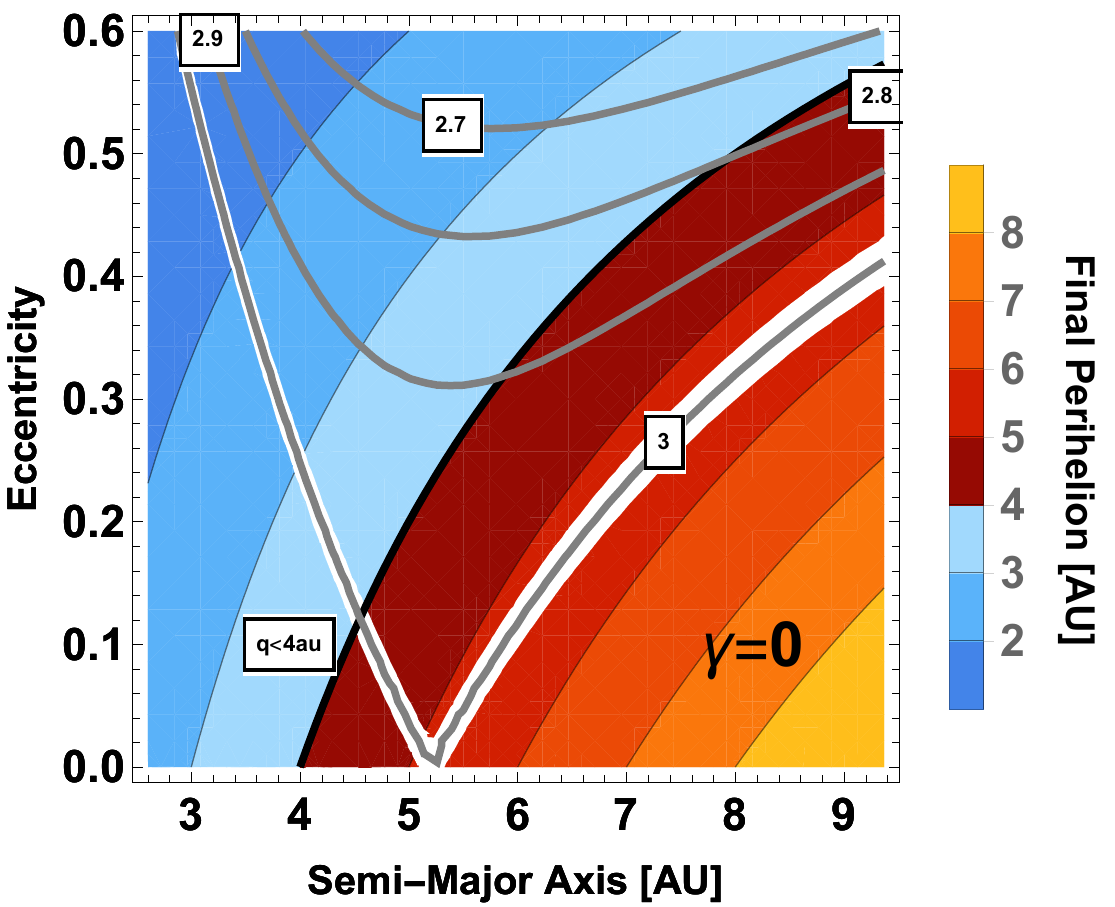}
\includegraphics[scale=0.7,angle=0]{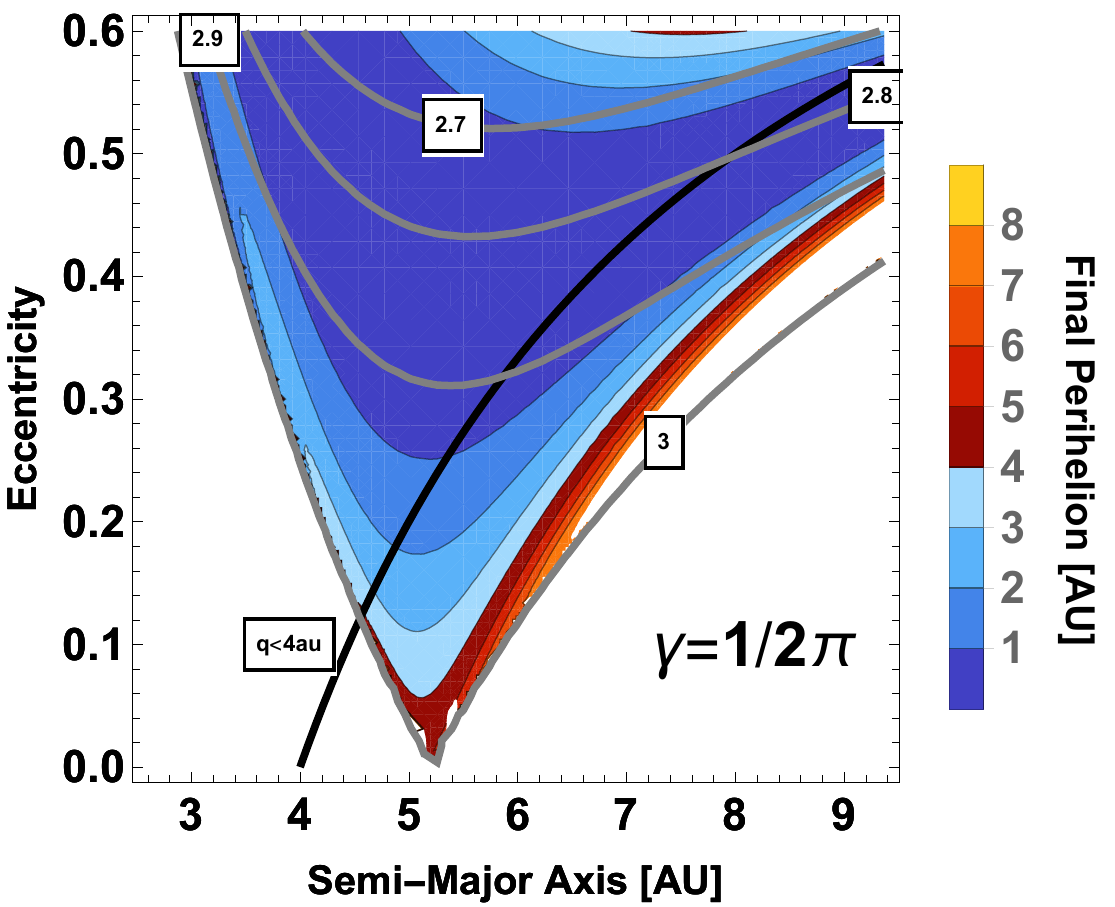}
\includegraphics[scale=0.7,angle=0]{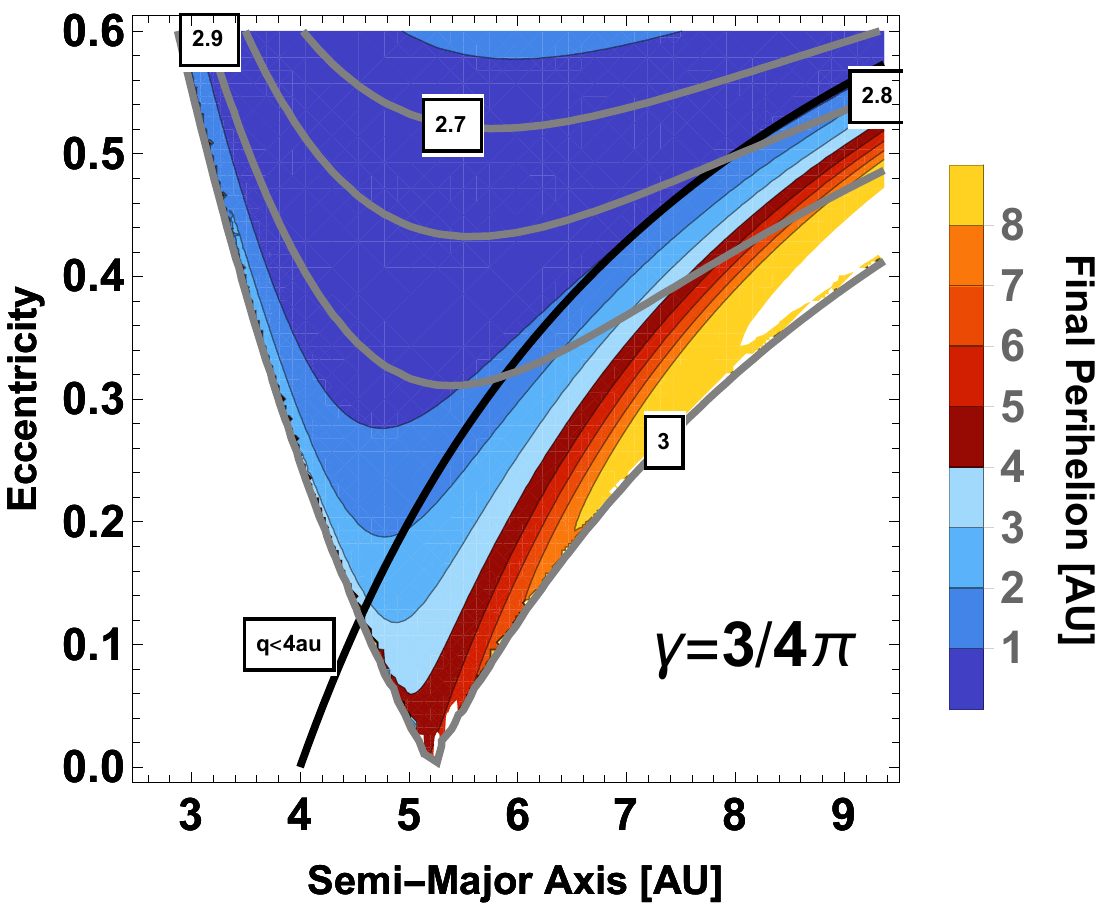}
\caption{The perihelion of the post-encounter orbit resulting from a  gravitational scattering event with Jupiter for objects with zero inclination. The final perihelion  of the resulting orbit  is color-coded and we show it for a range of initial semi-major axis and eccentricity. The black line shows where the initial perihelion is within $q<4$au. Contours of constant Tisserand parameter are plotted in grey lines and labeled. The three panels correspond to scattering angles, $\gamma$ of $0,\pi/2$ and $3\pi/4$.}\label{Fig:encounter}
\end{center}
\end{figure}

\begin{figure}
\begin{center}
\includegraphics[scale=0.45,angle=0]{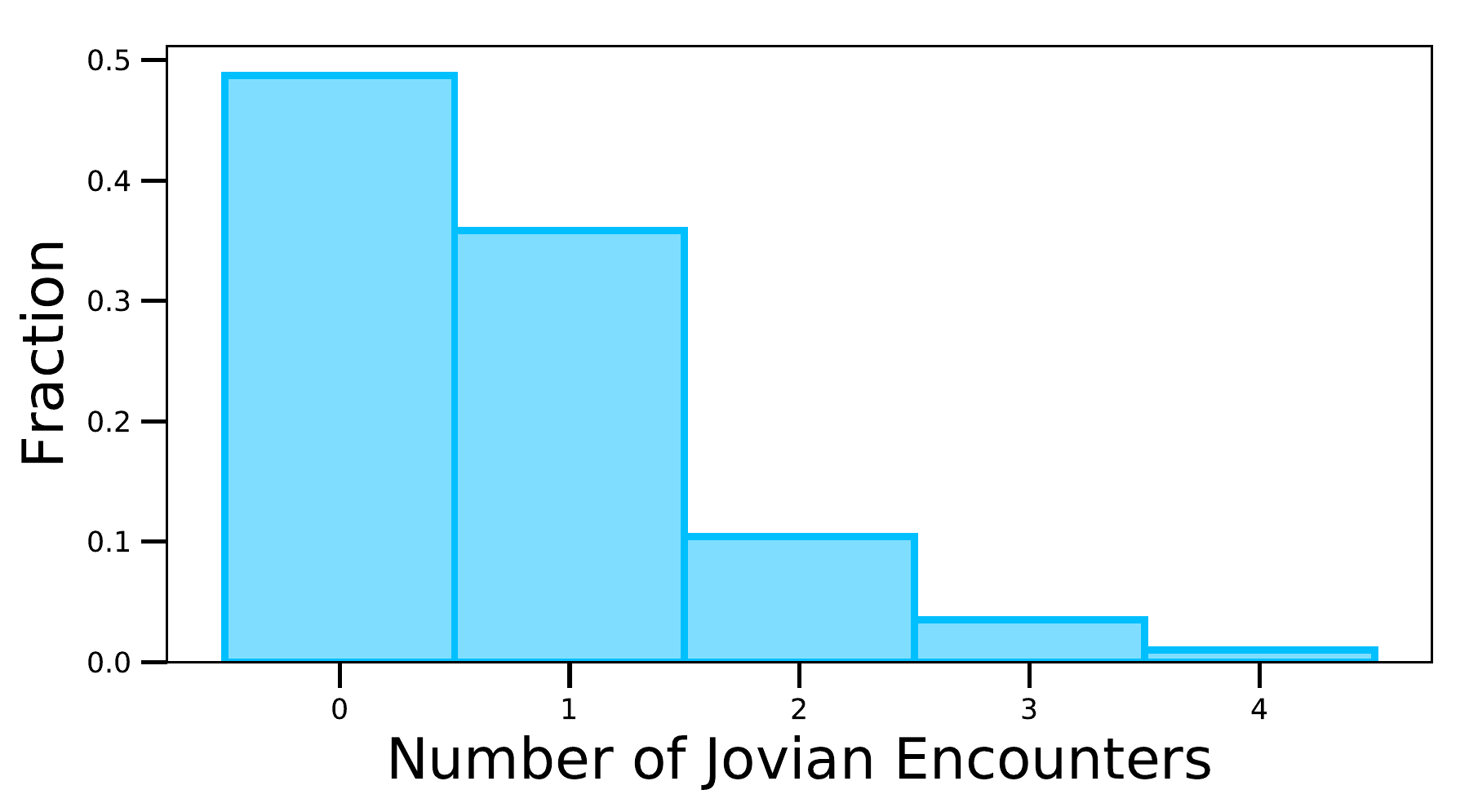}
\caption{The  number of Jovian encounters  before   simulated test particles  reach $q<4$au. A Jovian encounter is defined as an event  where the distance between the test particle and Jupiter is less than Jupiter's hill radius.}\label{Fig:number_scatters}
\end{center}
\end{figure}
\noindent where $M_J$ is the mass of Jupiter. In the frame co-rotating with Jupiter, the kinetic energy is conserved, so $|{\bf U}| =|{\bf U'}|$. The difference in the semi-major axis before and after the encounter, $\Delta a$, is given by Equation 13 in \citet{Carusi1990},
\begin{equation}\label{eq:delta_a}
   \Delta a=\frac{a'-a}{a}=\frac{1-|{\bf U}|^2-2|{\bf U}|\cos\theta'}{1-|{\bf U}|^2-2|{\bf U}|\cos\theta}-1\,.
\end{equation}
Here, $\theta'$ is determined by the \textit{unspecified} angle $\psi$, as given by Equation 11 in \citet{Carusi1990},
\begin{equation}\label{eq:thetap}
   \cos\theta'=\cos\theta\cos\gamma+\sin\theta\sin\gamma\cos\psi\,.
\end{equation}
Substituting the components of $|{\bf U}|$ into Equation \ref{eq:delta_a} yields a closed form solution for $a'$,
\begin{equation}\label{eq:ap}
\begin{split}
    a'=a-2a^2+\bigg(2a^4\sqrt{1-e^2}\bigg)\cos{i}\\-\bigg(2\sqrt{3a-1-2a^3\sqrt{1-e^2}\cos{i}}\bigg)\cos{\theta'}\,.
\end{split}
\end{equation}

The final eccentricity can be calculated using Equation 32 in \citet{Carusi1990},

\begin{equation}\label{eq:eccentricity}
   1-e'^2=\big(1-|{\bf U}|^2-2U_y'\big)\big(U_z'^2+(1+U_y')^2\big)\,.
\end{equation}
Unfortunately, it is not possible to derive closed form analytic solutions for the variation in eccentricity using this formalism. However, if we restrict our analysis to the case of zero inclination, $i=0$, where $U_z'=0$ then a closed form solution for the post-encounter eccentricity exists. Under these assumptions, Equation \ref{eq:eccentricity} can be written as,

\begin{equation}\label{eq:ep}
\begin{split}
    e'^2=1+\frac{1}{a}\bigg(1+\sqrt{3-1/a-2a^2\sqrt{1-e^2}}\cos{\theta'}\bigg)^2\\
    \bigg(-1+2a-2a^3\sqrt{1-e^2}+\\2\sqrt{3a^2-a-2a^4\sqrt{1-e^2}}\cos{\theta'}\bigg)\,.
\end{split}
\end{equation}

We solve these equations for three  deflection angles, $\gamma=\pi/4,\pi/2$ and $3\pi/4$, with $\psi=0$, and show the perihelion value of the post encounter orbit with Jupiter in Figure \ref{Fig:encounter}.  A single close encounter of an object with perihelia close to   Jupiter is capable of scattering it onto an orbit with $q<4$au. These results are consistent with the initial conditions for objects that reach $q<4$au in our simulations presented in Section \ref{sec:fullcentaur}. Since the kinetic energy of the object is conserved in the co-rotating frame, the gravitational scattering event cannot change the relative $\Delta$V with respect to Jupiter. Figure \ref{Fig:hypthothetical_mission} shows that nearly all of the objects scattered by Jupiter into the inner Solar System experience a period of low $\Delta$V with respect to the planet after their encounter. This dynamical feature merits further investigation. 

\subsection{Specific Case of LD2}

In the following section we show that LD2 reaches $\Delta$V $\sim 0.18$km/s in less than 2 years after its encounter with Jupiter. Moreover, it appears to be following a common evolutionary pathway from the outer Solar System into the inner Solar System. Presumably, the object migrated to the Centaur population from trans-Neptunian space. \citet{Steckloff2020} demonstrated that LD2 was likely in the  Gateway region recently, where MMRs with Jupiter and Saturn scattered it onto its current orbit. Besides the aforementioned close Jovian encounter, LD2 will have a second scattering event in 40 years. These events manifest themselves in Figure \ref{Fig:LD2_trajectory} as the two points where the distance between LD2 and Jupiter approaches zero.  Moreover, prior to the second scattering event, LD2 has $a\sim$7au and $e\sim0.3$, which places it in the region of Figure \ref{Fig:encounter} where objects can reach $q<4$au post-encounter for a range of impact parameters and resulting deflection angles $\gamma$, which serves as a validation of  the analytic methodology presented in the previous subsection.

This sequence of events could plausibly be representative of typical Centaur evolution. To test this hypothesis in our numerical simulations, we calculated the number of close encounters with Jupiter -- defined as occurring when objects enter Jupiter's Hill Sphere -- that test particles experienced prior to reaching $q<q$au. Figure \ref{Fig:number_scatters} shows that about $\sim45\%$ and $\sim35\%$ of the simulated objects that reach $q<4$au have 0 and 1 close encounters, respectively. The remaining $\sim20\%$ of objects experience multiple scattering events like LD2.

Jupiter can efficiently transfer objects into the inner Solar System via MMRs alone or via one close encounter. Moreover, we verified that the distributions of minimum perihelia attained and minimum $\Delta$V with respect to Jupiter are independent of the number of close encounters. Therefore, we conclude that LD2's dynamical evolution is representative of many objects that become SPCs. However, a close encounter with Jupiter is not required to produce these objects, and it is likely that the MMRs are efficient at scattering objects into the inner Solar System.

\section{Potential Mission to A Comet at the Onset of Intense Activity}\label{sec:missions}

\begin{figure}
\begin{center}
\includegraphics[scale=0.35,angle=0]{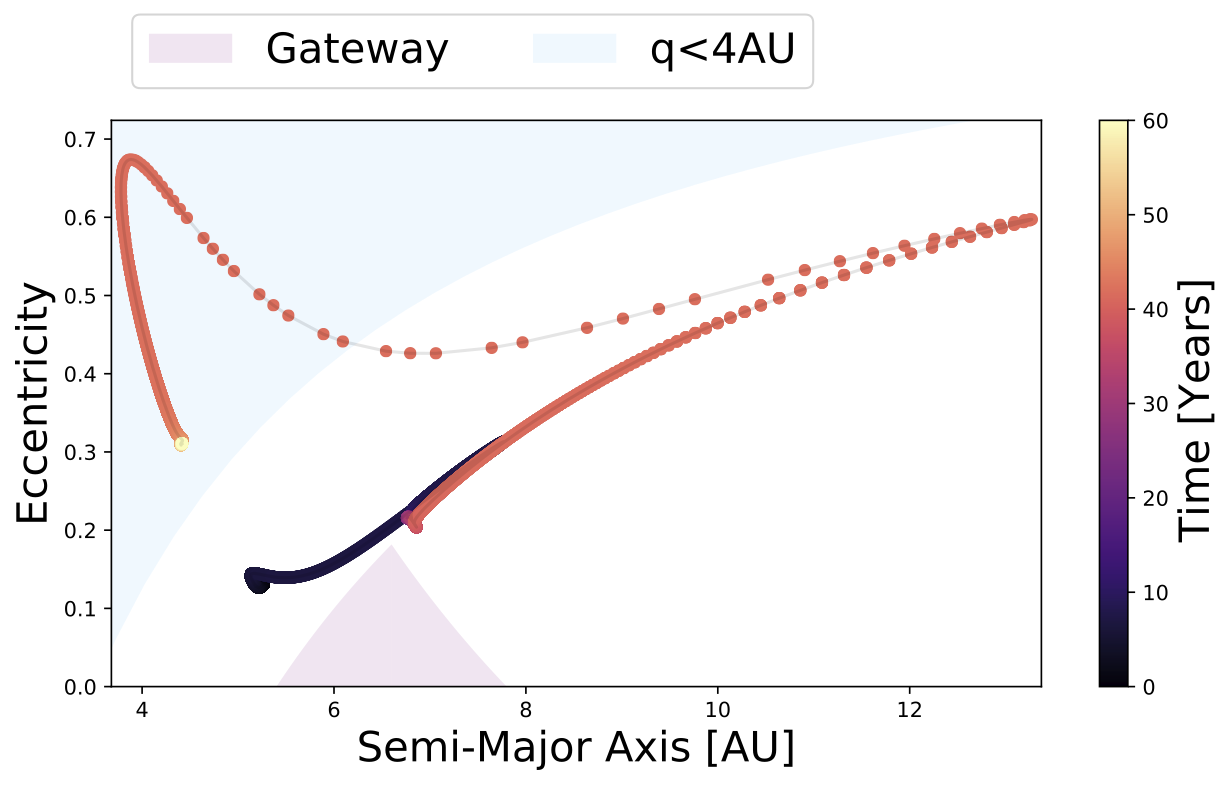}
\caption{The orbital evolution of LD2 in semi-major axis and eccentricity space. LD2's evolution is shown over the next 60 years, where the color of the points indicate time after 2021, and the points are plotted every  $\sim2$ days. The purple shaded region shows the nominal location of the Gateway, and the blue shaded region indicates where $q<4$au. LD2 experiences a close approach to Jupiter in 2063, which sends it into the inner Solar System quickly and dramatically changes the orbital elements. This figure may be directly compared to Figure 2 in \citet{Steckloff2020}.}\label{Fig:LD2_ae}
\end{center}
\end{figure}

LD2 represents an unprecedented opportunity to observe the evolution of cometary H$_2$O activity \textit{in situ} as it transitions into the inner Solar System. Such a study could unveil the evolution of surface features and the coma morphology during this transitory regime.  Here, we show that it will be viable to rendezvous with LD2 after the 2063 scattering event with Jupiter.

\begin{figure}
\begin{center}
\includegraphics[scale=0.35,angle=0]{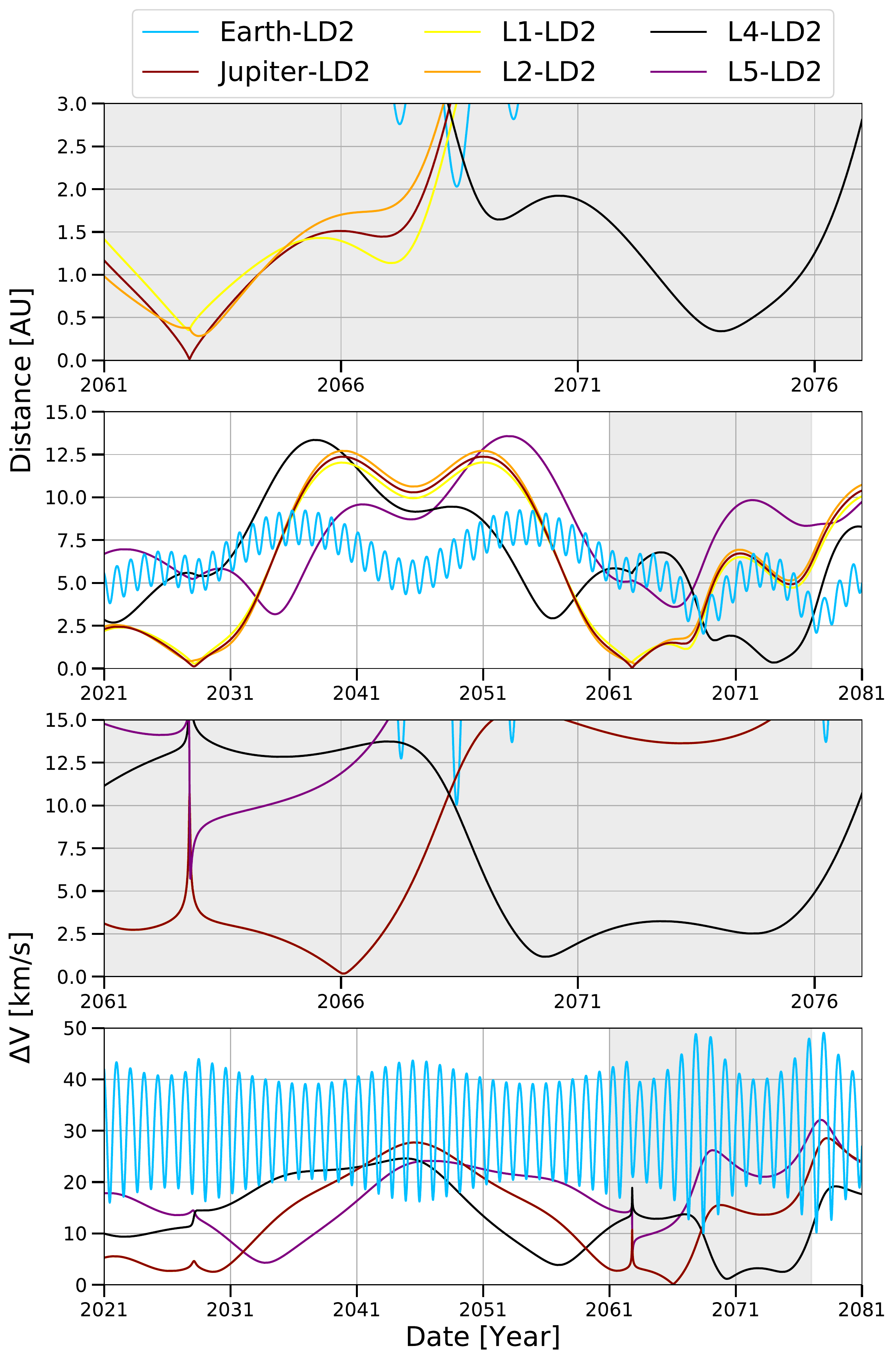}
\caption{The orbital separation and $\Delta$V of LD2 from selected starting locations in the Solar System in the next 60 years. The top two panels depict the orbital separation as a function of time between LD2 and the Earth, Jupiter and Jupiter's four Lagrange points, L1, L2, L4 and L5, in blue, red, yellow, orange, black and purple respectively. The bottom two panels show the difference in velocity between LD2 and each of these sites. The dates between 2061 and 2077 are shaded in grey.}\label{Fig:LD2_trajectory}
\end{center}
\end{figure}

\begin{figure*}
\begin{center}
\includegraphics[scale=0.7,angle=0]{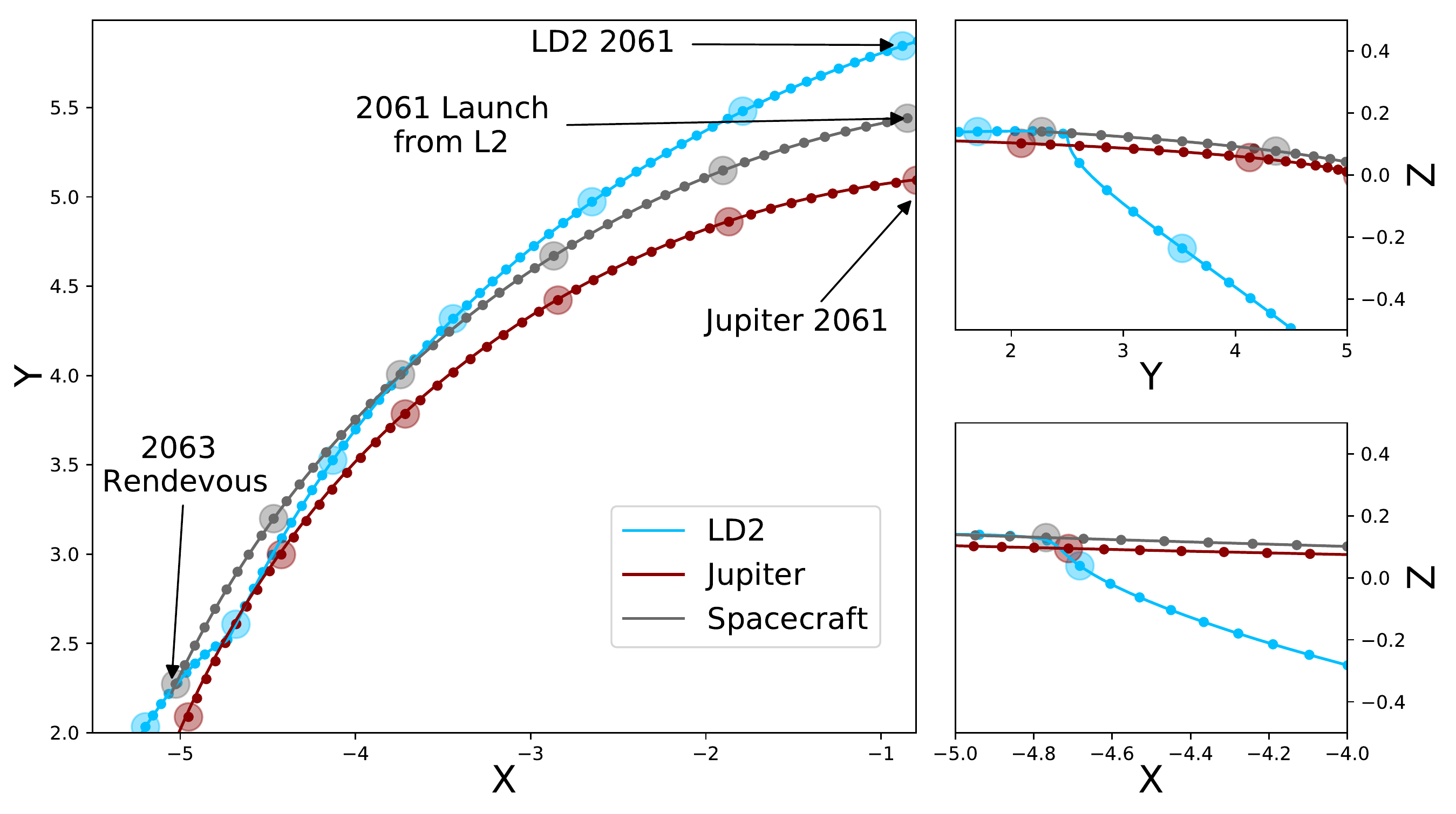}
\caption{A fiducial trajectory from the Jupiter-Sun L2 point to orbit match LD2. The orbits of LD2, Jupiter and the hypothetical spacecraft are shown in blue, red and grey respectively. Points along each trajectory correspond to evenly spaced sampling of the trajectory through time, where large and small circles correspond to two different cadences.  The spacecraft is sent in 2061 before LD2 experiences its closest approach to Jupiter, flies for $\sim2$ years, and rendezvous in 2063, after the close approach when the $\Delta$V between Jupiter and LD2 is small, in order to optimize the orbit matching efficiency. The required $\Delta$V from Jupiter's co-orbital location is $\sim0.9$km/s.}\label{Fig:LD2_orbits}
\end{center}
\end{figure*}

LD2 will undergo steady evolution through the JFC parameter space until it is scattered to large semi-major axis and eccentricity (Figure \ref{Fig:LD2_ae}). Then, it quickly returns to low semi-major axis and attains $e \sim 0.5$. This evolution of orbital elements slows as LD2 attains a perihelion distance of approximately $1.5$au.  In Figure \ref{Fig:LD2_ae}, we show the orbital evolution for LD2 calculated using the REBOUND N-body code \citep{rebound} with the hybrid symplectic MERCURIUS integrator \citep{reboundmercurius}. 

We investigate candidate points that could serve as stationary loitering locations for a spacecraft on a course to LD2. Specifically, Figure \ref{Fig:LD2_trajectory} shows the time evolution of LD2's orbital separation and instantaneous orbital $\Delta$V with respect to the Earth, Jupiter and the Jupiter-Sun Lagrange points L1, L2, L4 and L5. These curves were calculated by subtracting the distance and velocity of LD2 at each location in the numerical simulation described above. We note that this $\Delta$V is not the same as the $\Delta$V necessary to rendezvous. Over the next 60 years, LD2 will approach no closer than $\sim2$au to Earth, and its $\Delta$V with respect to Earth will never be below $\sim 10$km/s. It is important to note that although the orbital elements change drastically during the scattering event, the radial distance to the Sun does not change dramatically during this period.  While it is not infeasible to reach LD2 directly from Earth, a more promising alternative would be to park a spacecraft in the vicinity of Jupiter first. LD2 will travel within 0.02 au of the giant planet in 2063. Over the following two years, the relative $\Delta$V is $<2.5$ km/s. Therefore, the time period before 2063 is an ideal launch date from a Jovicentric orbit or either of the Jupiter-Sun L1 or L2 points to accompany LD2 as it first ventures into the inner Solar System. For a fiducial mission, we choose a launch date in 2061, to rendezvous shortly after the 2063 encounter. 

This type of mission concept is not unprecedented. The \textit{Juno} mission \citep{Bolton2017} reached Jupiter in less than 5 years, launched from Earth in 2011 on an Atlas V, with a launch energy per mass of 31.1 km$^2$/s$^2$, and using a combination of deep space maneuvers ($\Delta$V$\sim7.3$km/s) and orbit adjustments ($\Delta$V$<2$km/s),  reached  Jupiter in 2016 \citep{Kowalkowski2011}. The upcoming \textit{Lucy} mission will fly to the Trojan region at L4, and visit several Jupiter Trojan asteroids \citep{Olkin2021}. While the L4 and L5 points may be scientifically-appealing parking locations as an opportunity to visit Jupiter's Trojan asteroids, their relative $\Delta$V  would pose a challenge. Because these points form equilateral triangles with the Sun and Jupiter, they are located far from the close encounter between LD2 and Jupiter. The third panel of Figure \ref{Fig:LD2_trajectory} focuses on the time close to 2063 and shows that there is a timespan of $\sim5$ years where the relative $\Delta$V between LD2 and L4 is $<5$ km/s and the minimal orbital separation is approximately 0.5 au. %Although both L1 and L2 have close orbital separations and small relative $\Delta$V, parking at the former would permit continuous monitoring of the fully-illuminated Jovian atmosphere between arrival and departure for LD2.
 
 %While the Trojan and Greek regions seem appealing as parking locations, due to their stability and the additional scientific benefits of observations of these asteroids, the relative $\Delta$v and distances with LD2 from these locations, would pose a challenge.

With either L1 or L2 as a parking location and  2061 as a launch date, we may formulate the problem as finding the optimal trajectory for a spacecraft to rendezvous with LD2. The payload of the spacecraft would need to provide sufficient fuel to (1) reach L1 or L2 of Jupiter (comparable to $\Delta {\rm V}<2$km/s that \textit{Juno} used), (2) provide an impulsive thrust from the parking location to rendezvous with LD2  (to be determined), and (3) orbit match LD2 upon rendezvousing ($\Delta {\rm V}<2.5$km/s).  Given two vectors for both position ($\bf{r_0}'$, $\bf{r_0}$) and velocity ($\bf{v_0}'$, $\bf{v_0}$) for a spacecraft and target at initial time $t_0$, and given a predetermined optimal flight time $\Delta t$, we can solve for the impulsive change to the initial velocity ${\bf\Delta v_0}'$ which ensures a rendezvous orbit. For the case of two elliptical orbits, \citet{Leeghim2013} developed a robust algorithm to optimize  the trajectory using Lagrange multipliers. In this formulation, both target and interceptor orbits needed to be solved for at some time in the future, and the flight time could be optimized in order to minimize the kinetic energy necessary for an interception. However, for the case of LD2, we have dictated the flight time, and it is sufficient to specify the launch  and rendezvous dates. %Moreover, since LD2 is undergoing drastic changes in its orbital elements due to the close encounter with Jupiter close to the time when we aim to launch, it is simpler to simply integrate LD2's position using an N-Body simulation, and optimize our interception to the location that LD2 is under the influence of Jupiter's gravitational effect. 

Following \citet{Bate1971} and \citet{Chobotov1991}, it is useful to adopt the universal variable, $\chi$, which is defined as

\begin{equation}\label{eq:chi}
\chi =\sqrt{a}E  \, ,
\end{equation}
for an elliptical orbit. Here, the orbit's semi-major axis is $a$, and eccentric anomaly is $E$. Given a value of $\chi$ at a time $t_0+\Delta t$, Kepler's equation can be written,
\begin{equation}\label{eq:kepler}
    \sqrt{\mu}\Delta t = \frac{{\bf r_0\cdot v_0}}{\sqrt{\mu}}\chi^2C(\alpha \chi^2)+(1-\alpha r_0)\chi^3S(\alpha \chi^2)+r_0\chi\, ,
\end{equation}
where $\alpha = 1/a$, $\mu=GM_\odot$, and $C(x)$ and $S(x)$ are Stumpff functions, defined as

\begin{equation}\label{eq:stumpff1}
    S(x) =  \frac{\sqrt{x}-\sin\sqrt{x}}{\sqrt{x}^3}\, ,
\end{equation}
and
\begin{equation}\label{eq:stumpff2}
    C(x) = \frac{1-\cos\sqrt{x}}{\sqrt{x}}\, ,
\end{equation}
for the case of $x>0$. The four Lagrange coefficients, commonly referred to as dynamical $f$ and $g$ functions, are given by,

\begin{equation}\label{eq:f}
    f = 1-\frac{\chi^2}{r_0}C(\alpha\chi^2)\, ,
\end{equation}

\begin{equation}\label{eq:g}
    g = \Delta t -\frac{1}{\sqrt{\mu}}\chi^3S(\alpha\chi^2)\, ,
\end{equation}
\begin{equation}\label{eq:fdot}
    \dot{f}=\frac{\sqrt{\mu}}{rr_0}[\alpha \chi^3 S(\alpha\chi^2)-\chi]\, ,
\end{equation}
and
\begin{equation}\label{eq:gdot}
    \dot{g}=1-\frac{\chi^2}{r}C(\alpha\chi^2)\, .
\end{equation}
The universal variable and dynamical $f$ and $g$ functions are particularly useful for astrodynamics problems, as they uniquely determine the position and velocity vector of an orbit after a time $\Delta t$ via,

\begin{equation}\label{eq:position}
    {\bf r}(\Delta t)=f {\bf r_0}+g{\bf v_0} \, ,
\end{equation}
and
\begin{equation}\label{eq:velocity}
    {\bf v}(\Delta t)=\dot{f}{\bf r_0}+\dot{g}{\bf v_0}\, ,
\end{equation}
where ${\bf r}(\Delta t)$ and ${\bf v}(\Delta t)$ are the positions and velocity vectors after a time, $\Delta t$. It is important to note that Equations \ref{eq:f}- \ref{eq:gdot} are implicitly dependent only on $\Delta t$, except for the case of $g$ which is explicit, because $\chi$ is uniquely determined by a set of orbital elements and $\Delta t$. 
As defined by \citet{Leeghim2013}, we adopt the function, $\eta$, which expresses the orbital motion of the interceptor in terms of its  universal variables,

\begin{equation}\label{eq:eta2}
\begin{split}
\eta(\chi',{\bf \Delta v_0'}, \Delta t) = \frac{{\bf r_0'\cdot (v_0'+\Delta v_0')}}{\sqrt{\mu}}\chi'^2C(\alpha \chi'^2)\\+(1-\alpha r_0')\chi'^3S(\alpha \chi'^2)+r_0'\chi'-\sqrt{\mu}\Delta t \, .
\end{split}
\end{equation}
$\eta(\chi',{\bf \Delta v_0'})$ describes the full evolution of the spacecraft after a pre-determined flight time. It is particularly useful to cast Kepler's equation in this form, because to solve for real solutions to Kepler's equations corresponding to real orbits, $\eta(\chi',{\bf \Delta v_0'})=0$. Therefore, the problem amounts to finding values, $\bf{\Delta v_0'}$, for a given $\Delta t$, that are roots of the transcendental equations defined by,
 
 \begin{equation}\label{eq:equalpos}
     {\bf r'}({\bf \Delta v_0'}, \Delta t) - {\bf r}( \Delta t)=0\, ,
 \end{equation}
 and
 \begin{equation}\label{eq:eta20}
     \eta(\chi',{\bf \Delta v_0'}, \Delta t)=0\, .
 \end{equation}
The constraints in Equation \ref{eq:equalpos} demand that the positions of the orbits are equal after the designated flight time. The second constraint, Equation \ref{eq:eta20},  demands that the orbit of the interceptor after the single, impulsive change in velocity satisfies Kepler's equations.  Numerical solutions for this system of equations are straightforward to identify following an iterative process to solve the system of four transcendental equations defined by Equations \ref{eq:eta2} and \ref{eq:equalpos} for all three components of ${\bf \Delta v_0'}$ and $\chi'$.

 %In Figure \ref{Fig:LD2_trajectory}, we show the orbital evolution of LD2 over the next 60 years. In the top two panel, we show the seperation between LD2 and the Earth, Jupiter and the four Lagrange points that could be stationary parking locations for a rendevous mission. In the bottom two panels, we show the difference in velocity between LD2 and each of these locations as a function of time. It is evident that in 42 years in 2063, LD2 will experience a very close approach to jupiter and come within $0.1AU$ of the giant planet. This poses an optimal launch time for a rendevous target. Just two years later in 2065, the delta v between Jupiter and its L1 and L2 and LD2 is very small and approaches almost zero. In the span of time bewteen 2063 and 2065, the delta v between Jupiter and LD2 is less than $<2.5$ km/s. Therefore, 2063 is an ideal target date for a launch from a geosynchronous orbit with Jupiter or either of the Lagrange points, to reach LD2 in 2065 as it begins its descent into the inner Solar System.

For the case of LD2, we iterate for the optimal solution that guarantees that the interceptor matches the position of LD2 when launched from L2 in 2061, to guarantee a rendezvous in 2063. The optimization criteria was defined as a solution where the residuals of Equations \ref{eq:equalpos} and \ref{eq:eta20} added in quadrature were smaller than $10^{-3}$ in magnitude. We numerically verified that this criteria was computationally efficient while still maintaining a difference in location upon rendezvous of $<10^{-5}$au.  We calculate  an optimal solution that requires a $\Delta {\rm V}\sim0.93$ km/s.  The components of the velocity are $\Delta {\rm V_x}=0.28$, $\Delta {\rm V_y}=-0.87$, and $\Delta {\rm V_z}=0.10$ km/s. The same procedure launched from Jupiter's position yields a $\Delta {\rm V}\sim0.4$ km/s, but this does not include the $\Delta$V required to escape from Jupiter's  gravitational potential. The total $\Delta$V of the 3 phases is conservatively $\Delta {\rm V}\lesssim6$ km/s, which would be attainable with a Falcon Heavy or Atlas V \citep{Seligman2018}.

Figure \ref{Fig:LD2_orbits} shows the orbits of LD2, Jupiter and the hypothetical spacecraft.  The close approach to Jupiter reorients LD2 into the ecliptic plane, making the rendezvous and required $\Delta$V  attainable. As can be seen in the left panel in the x-y plane at $\sim(-4.8,2.6)$au, the close approach to Jupiter in 2063 appears as a kink in LD2's orbit. The change in the trajectory is much more dramatic when viewed from the x-z (at $\sim(-4.7,0.0)$au) and y-z plane (at $\sim(2.5,0.0)$au)(not scaled to equality), in the right two panels. While LD2 approaches Jupiter from below the ecliptic, Jupiter re-orients  the orbital angular momentum vector so that it is almost perpendicular to Jupiter's orbital plane. This amounts to to a small required $\Delta$V in the z-direction for our hypothetical rendezvous spacecraft.   Once this rendezvous occurs in 2063 at $\sim(-5.1,2.3)$au, shortly after the close approach, the spacecraft would have to re-adjust its orbital velocity with a $\Delta {\rm V}\sim2.5$ km/s in order to orbit match LD2 as it begins its journey into the inner Solar System. The exact dates of launch and rendezvous can be altered to optimize the  $\Delta$V requirement. We chose 2061 and 2063 as a fiducial launch and rendezvous dates, to minimize the distance that the spacecraft needs to travel, while still reaching LD2 long enough after the close encounter to have a feasible $\Delta$V required to orbit match.

It is important to note the uncertainty associated with the evolution of LD2's orbit, as demonstrated  by \citet{Steckloff2020}. Due to the chaotic nature of orbits in this region of the Solar System, the uncertainties of the osculating elements of LD2 could lead to a diversity of potential outcomes for the object. Moreover, cometary activity-driven non-gravitational forces should alter the trajectory before the 2063 scattering event, especially if the nucleus's size is towards the smaller end of estimates. \citet{Steckloff2020} simulated 1000 model clones of LD2 by sampling the JPL orbit fit covariance matrix from May 2020. They found that the orbital histories tended to diverge in backward integrations before $\sim 1770$. Indeed, the orbit that we calculate using the JPL  fit from June 2021 is different from the one presented in \citet{Steckloff2020}. This alone is a good indicator that there is significant uncertainty in the long term evolution of LD2.

Nonetheless, the close encounter with Jupiter in 2063 has a $>98\%$ probability of scattering LD2 into the inner Solar System, making LD2 a worthwhile target for a mission \citep{Kareta2020,Hsieh2021,Steckloff2020}. From the encounter simulations described above, the most likely scenarios would all exhibit low $\Delta$V with respect to Jupiter, making this mission concept feasible, despite the uncertainties. In the upcoming decades, follow-up observations of LD2 will decrease the uncertainty in its trajectory, and allow for  monitoring of changes in the trajectory due to non-gravitational forces.  Moreover, as demonstrated in Section \ref{sec:fullcentaur}, all of the pathways that generate an object with $q<4$au, \textit{including} those that do not experience a close encounter with Jupiter, exhibit a period of low $\Delta$V with respect to Jupiter. Therefore, if additional transitioning objects are detected, they should all be attainable targets for this type of rendezvous.

\section{Discussion}\label{sec:discussion}

\subsection{Other Objects Like LD2}

LD2 represents the best known opportunity to monitor  the onset of intense cometary  activity within $q<4$au in a pristine small body \citep{Steckloff2020}. In \S \ref{sec:missions}, we show that a spacecraft stationed at the Jupiter-Sun Lagrange point with reasonable $\Delta\mathrm{V}\lesssim1$km/s could rendezvous with LD2 in 2063, and that orbit matching the object would be attainable with an additional $\Delta$V of $\sim 3$ km/s.

\citet{Sarid2019} demonstrated that objects that transitioned from the Centaur population  to the JFCs passed through  a  Gateway of low eccentricity orbits close to Jupiter. They found that roughly half of the objects that reach  $q<4$au occupy this Gateway prior to transitioning in their numerical simulations. In \S\ref{sec:dynamics}  and \S\ref{sec:simulations} we explored the impact of MMRs with Jupiter and Saturn in this Gateway region.  Based on the estimated occupancy of the Gateway region from \citet{Sarid2019} and \citet{Steckloff2020}, we estimated that $\sim1-2$ (with large uncertainty) additional objects from this region could become comets that reach $q<4$au within the next 50 years. From our simulated population of JFCs and Centaurs, in \S \ref{sec:fullcentaur} and \S \ref{sec:scattering}, we show that if additional targets transition, they will also exhibit low $\Delta$V with respect to Jupiter, and should be feasible targets for a rendezvous.

\subsection{Future Observational Constraints on Imminently-Active Objects}

The forthcoming Rubin Observatory Legacy Survey of Space and Time (LSST)  may increase the number of known minor bodies in the Solar System by a factor of 25 \citep{Ivezic2019}. It will provide unprecedented completeness for both asteroidal and cometary populations \citep{jones2009lsst}. Furthermore, the ability of the LSST to detect transient objects has already been demonstrated for Near Earth Objects \citep{Veres2017,veres2017b,Jones2018}, and it will efficiently detect both JFCs and LPCs changing between active and inactive states \citep{solontoi2011comet}. Because the Centaurs contain similar volatile profiles and also make this transition to higher sublimation rates, we expect LSST to efficiently detect these objects as well.

If other Centaurs identified by VRO/LSST other than LD2 are found to be imminently transitioning into the inner Solar System, follow-up observations and orbit determinations may allow us to plan an optimal trajectory mission to rendevous and orbit match with those objects as well. Tighter constraints and confirmation of the population size and distribution of orbital elements for Centaurs from this survey will permit more detailed predictions for the number of pristine objects that transition to the inner Solar System  on human timescales.

\section{Acknowledgements}
We thank Kat Volk, Renu Malhotra, Adina Feinstein, Konstantin Batygin, Juliette Becker, Andrew Youdin,  Megan Mansfield, Sam Cabot, Marvin Morgan, Gal Sarid, Daniel Fabrycky, and Zachary Claytor for useful conversations. We thank the scientific editor, Maria Womack, and the two anonymous reviewers for insightful comments and constructive suggestions that strengthened the scientific content of this manuscript.
\bibliography{sample63}{}
\bibliographystyle{aasjournal}



\end{document}